\documentclass[11pt,english]{article}
\usepackage[T1]{fontenc}
\usepackage[latin9]{inputenc}
\usepackage{geometry}
\usepackage{color}
\usepackage{babel}
\usepackage{float}
\usepackage{amsmath}
\usepackage{amssymb}
\usepackage{graphicx}
\usepackage{setspace}
\usepackage[unicode=true]
 {hyperref}

\makeatletter

\providecommand{\tabularnewline}{\\}
\floatstyle{ruled}
\newfloat{algorithm}{tbp}{loa}
\providecommand{\algorithmname}{Algorithm}
\floatname{algorithm}{\protect\algorithmname}

\usepackage{babel}
\date{}

\makeatother

\begin{document}

\title{Signals as Parametric Curves: Application to Independent Component
Analysis and Blind Source Separation}

\author{Birmingham Hang Guan and Anand Rangarajan\thanks{The authors are with the Department of Computer and Information Science
and Engineering, University of Florida, Gainesville, FL, USA. E-mail:
\protect\href{mailto:{bkwan,anandr}@ufl.edu}{{bkwan,anandr}@ufl.edu}.}}
\maketitle
\begin{abstract}
Images Stacks as Parametric Surfaces (ISPS) is a powerful model that
was originally proposed for image registration. Being closely related
to mutual information (MI) \textendash{} the most classic similarity
measure for image registration, ISPS works well across different categories
of registration problems. The Signals as Parametric Curves (SPC) model
is derived from ISPS extended to 1-dimensional signals. Blind Source
Separation (BSS) is a classic problem in signal processing, where
Independent Component Analysis (ICA) based approaches are popular
and effective. Since MI plays an important role in ICA, based on the
close relationship with MI, we apply SPC model to BSS in this paper,
and propose a group of geometrical objective functions that are simple
yet powerful, and serve as replacements of original MI-based objective
functions. Motivated by the geometrical objective functions, we also
propose a second-order-statistics approach, FT-PCA. Both geometrical
objective functions and FT-PCA consider signals as functions instead
of stochastic processes, make use of derivative information of signals,
and do not rely on the independence assumption. In this paper, we
discuss the reasonability of the assumptions of geometrical objective
functions and FT-PCA, and show their effectiveness by synthetic experiments,
comparing with other previous classic approaches.
\end{abstract}

\section{Introduction\label{sec:Introduction}}

Independent component analysis (ICA) (\cite{comon2010handbook,hyvarinen2001independent})
is a well-known topic in machine learning, statistics, and signal
processing. The original ICA problem (\cite{comon1994independent,bell1995information})
and its various extensions (\cite{hyvarinen1999fast,bach2002kernel})
have been researched through the past 25 years. Being a theoretical
topic in statistics, it was originally proposed and applied for signal
processing problems, especially blind source separation (BSS) (\cite{comon1994independent}).
ICA-based BSS has various practical applications, like electroencephalographic
data analysis (EEG) (\cite{delorme2004eeglab,makeig1996independent}).
ICA was also applied to image problems, like the work in \cite{lee2002unsupervised,mitianoudis2007pixel},
where the latent independent variable linear mixture model were used
for image fusion, etc..

Formally, and ICA problem is described in the following manner: Let
$\mathbf{x}$ be a random vector of size $N$. The generative model
for$\mathbf{x}$ is via a matrix $A$ of size $N\times M$ and a random
vector $\mathbf{s}$ of size $M$ whose components are independent
of each other, such that 
\[
\mathbf{x}=A\mathbf{s}.
\]
Mutual information (MI), as a natural independence measure of random
variables, is considered as a standard approach to solve for the ICA
problem:
\[
MI(\mathbf{s})=\int p_{\mathbf{s}}(\mathbf{s})\log\frac{p_{\mathbf{s}}(\mathbf{s})}{\prod_{i}p_{s_{i}}(s_{i})}\mathrm{d}\mathbf{s}.
\]
If each $s_{i}$ are independent, $MI(\mathbf{s})=0$. Therefore,
by minimizing MI, one can get separated random variables as independent
as possible.

The definition of a BSS problem has a very close structure: Let $\mathbf{x}(t)=(x_{1}(t),\dots,x_{N}(t))$
be a set of observed signals. Supposing that they are a linear mixture
of a set of unknown source signals $\mathbf{s}(t)=(s_{1}(t),\dots,s_{M}(t))$,
we can express it as 
\[
\mathbf{x}(t)=A\mathbf{s}(t)
\]
where $A$ is the mixing matrix. There are two main differences between
an ICA problem and a BSS problem: First, each component of the $\mathbf{x}$
and $\mathbf{s}$ in the ICA problem is a random variable, whose density
is estimated by observations of the sample data; while the components
of $\mathbf{x}(t)$ and $\mathbf{s}(t)$ are signals, which can be
considered as stochastic processes. The core difference is about the
data being discrete or continuous. Second, being random variables,
the components of $\mathbf{s}$ in the ICA problem are independent.
In BSS, independence is not a necessary assumption. 

Apparently, BSS is a highly open and nondeterministic problem, where
both $A$ and $\mathbf{s}(t)$ need to be determined. For any approach
to solve BSS, additional assumption on sources should be made to decrease
the infinitely many solutions down to a small subset. Similar to ICA,
the signals $\mathbf{x}(t)$ and $\mathbf{s}(t)$ are commonly assumed
to be zero-mean, and signals in $\mathbf{s}(t)$ are uncorrelated
with each other. Under this assumption, most approaches applied principal
component analysis (PCA) to standardize signals as the first step.
As early as 1981 (\cite{mermoz1981spatial}), it was pointed out that
the information in the spectral matrix is not sufficient for separation,
and additional assumptions are needed. The work in \cite{lacoume1988sources,comon1989separation}
started to consider higher order statistics, and the assumption of
source signals being independence was put forth (\cite{comon1989separation,comon1991blind}).
From then on, the assumption of independence was taken for granted
by most approaches (and these are not necessarily restricted to just
ICA) (\cite{belouchrani1997blind,cardoso1998blind}) for BSS.

Though the independence assumption in ICA is natural, few papers pointed
out the difference in models by applying ICA to BSS. The independence
of a set of signals is defined by the independence of the distributions
underlying each signal, considering each signal as a sample function
from a stationary stochastic process whose distribution at each time
point is identical. By the ergodicity of the stationary stochastic
processes, the sample values of the signals can be used to estimate
the statistical properties of the distributions, and hence can be
used to estimate the MI and independence. This is different from ICA,
where data are observations of samples of each underlying distribution.
Unfortunately, most ICA work, like the work in \cite{hyvarinen2000independent,comon1994independent}
just mention that consideration of time $t$ should be neglected,
and the signal values should be considered as a collection of unordered
observations. The drawback of this perspective is that information
contained in the ``changing over time'' of signals are ignored. 

Another perspective is to consider time $t$ as a uniform distributed
random variable, like the work in \cite{rajwade2009probability}.
In this way, being continuous functions of time, signals can still
be considered as random variables and the ``changing over time''
information of signals are taken into consideration. However, under
this model, independence does not exist since all signals are functions
of the same random variable. Therefore, the definition of independence
of signals is not well-defined under this perspective. 

For BSS, the independence assumption is not necessary. Similar to
the work in \cite{belouchrani1997blind,belouchrani1998blind}, we
are dedicated to consider additional information provided by signals
as functions, and solve the BSS problems without the assumption of
independence. Signals can be either considered as deterministic functions
of time, or stochastic processes, based on its application. Considering
signals as deterministic functions, we extend the Images Stacks as
Parametric Surfaces model (ISPS), a powerful model originally designed
for image registration, to 1D case (we call it Signals as Parametric
Curves (SPC), accordingly), and apply SPC to BSS. Based on the close
relationship between SPC and MI, we propose geometrical objective
functions that can approximate the MI-based objective functions. We
are also able to analyze signals in the frequency domain by the Fourier
transform of the signals, and propose the FT-PCA algorithm, which
does not rely on the independence of signals, and focuses on the local
orthogonality in the frequency domain. For simplicity, in this paper,
we only focus on the two-dimensional case (where there are only two
observed signals and two source signals), and higher dimensional cases
can be extended naturally. 

The main content of this paper is as follows: Section \ref{sec:Previous-Work}
briefly introduces well-known previous approaches for BSS, including
ICA and second-order-statistics approaches; Section \ref{sec:ICArevis}
briefly summarizes and analyzes the ICA framework and its assumption;
Section \ref{sec:Applying-SPC-to} applies SPC to BSS and propose
geometrical objective functions that are competitive with the traditional
MI approach; Section \ref{sec:FT-PCA} introduces FT-PCA algorithm
based on the assumption of kernel-orthogonality in the frequency domain;
Section \ref{sec:6-Experiments} shows synthetic simulation experiments
and compare our approaches with several well-known approaches, and
shows the effectiveness of our algorithms; and the paper is concluded
in Section \ref{sec:Conclusions} to highlight our simple yet effective
approaches for BSS.

\section{Previous Work\label{sec:Previous-Work}}

Most approaches of BSS can roughly be categorized into two classes:
high-order statistics based approaches, or second-order statistics
based approaches. ICA approaches stick to the assumption of independence,
and try to minimize the entropies of signals to recover sources which
are as independent as possible; while joint diagonalization approaches
try to make use of information and properties of second-order statistics
of signals to solve for the unmixing matrix, bypassing the direct
usage of independence to avoid higher-order statistics.

The work in \cite{comon1989separation,comon1991blind,comon1994independent}
firstly introduced the concept of ICA, and created the independence
assumption as a foundation of BSS. \cite{bell1995information} is
another well-known paper that highlighted mutual information based
approaches for BSS. In the original ICA framework, the objective function
was directly based on the assumption of independence: the Kullback
divergence of the joint density and the product of marginal densities,
i.e. the mutual information. Nevertheless, its estimation is difficult,
and high-order cumulants were introduced to estimate entropies. The
work also suggested a standardization step using PCA to standardize
the deviation, and pointed out that after PCA, the minimization of
MI is equivalent to minimization of negentropies with respect to a
sequence of pairwise rotations of signals. The work in \cite{hyvarinen1999fast}
put forward the FastICA algorithm. Based on their previous work in
estimating entropies (\cite{hyvarinen1998new}), they suggested a
set of contrast functions that are much simpler to compute than high
order cumulants in the work in \cite{comon1994independent}. They
also adopted Newton method to decrease the time complexity, so that
each signal can be optimized one by one. The FastICA algorithm is
very efficient and widely used until now. Another most well-known
approach in the ICA category is Kernel ICA, proposed in the work in
\cite{bach2002kernel}. The goal of Kernel ICA is to maximize the
kernel correlation of whitened signals. It constructs an eigen-decomposition
structure, and computes the minimum eigenvalue of a matrix constructed
by certain Gram matrices of signal data points. Though the idea is
somehow close to our kernel orthogonality, the approach is totally
different. It is still within the ICA optimization framework.

Comparing to the ICA series where signals values are used to estimate
the independence of the underlying distributions, the second-order-statistics
class (we call it SOBI series) tried to take use of other stochastic
process properties to bypass the approximation of entropies. AMUSE
(\cite{tong1990amuse}) algorithm is an early work of these approaches.
Its assumption on source signals is that given some time shift $\tau$,
the auto-correlation matrix is diagonal but not identity, i.e. for
$i\ne j$, $E(s_{i}(t)s_{i}(t-\tau))\ne E(s_{j}(t)s_{j}(t-\tau))$
and $E(s_{i}(t)s_{j}(t-\tau))=0$. This assumption grants another
eigen-decomposition structure than the PCA step, and makes AMUSE an
approach where no optimization is required. However, not all $\tau$
grants diagonal matrices. Once the selected $\tau$ makes the auto-correlation
matrix isomorphic to identity, the eigen-decomposition gives trivial
results, and AMUSE fails. The work in \cite{belouchrani1997blind}
put forward a joint diagonalization scheme, and an extended algorithm,
named SOBI. Instead of a certain $\tau$, SOBI is based on the assumption
that $E_{\tau}(\mathbf{s}(t+\tau)\mathbf{s}^{H}(t))$ is diagonal,
assuming that $\mathbf{s}$ is a multivariate stationary process of
both $t$ and $\tau$. (It also has an equivalent assumption where
the expectation of $\tau$ is defined as arithmetic average of a set
of different $\tau$'s.) To select a bunch of different $\tau$ and
use the joint diagonalization scheme, SOBI avoided the occurrence
of a single trivial $\tau$, and is able to solve the problem by $K$
times matrix diagonalization, where $K$ is the number of $\tau$
selected. A following work in \cite{belouchrani1998blind} extended
this idea to non-stationary signals, where time-frequency distribution
(TFD) (\cite{cohen1995time}) was introduced. Based on similar fact
that the spatial TFD matrices (STFD) of signals being diagonal but
not identity, eigen-decomposition scheme is also able to be applied
to STFD matrices. Since STFD are dependent with time and frequency
indices $(t,f)$, and for some special $(t,f)$, the STFD matrix can
be rank deficient, they again applied joint diagonalization scheme
to solve the problem by a set of different selected $(t,f)$. STFD
is close to our approach, except that designed for non-stationary
signals, the time-frequency domain analysis was introduced. And similar
to SOBI, it adopted selection of parameters and joint diagonalization.
Though this approach can handle non-stationary signals and Gaussian
signals, it was criticized by complexity and performance (\cite{abrard2005time,james2004independent}).
After this work, many following work came out based on time-frequency
analysis and joint diagonalization (\cite{chabriel2014joint}). However,
most of them, like the work in \cite{fevotte2004two}, did not improve
the fact that STFD needs local parameter selection and joint diagonalization,
and focus on non-stationary signals, which is out of the scope of
this paper.

The work in \cite{jourjine2000blind} is another one close to ours.
It also put forward the assumption of disjoint orthogonality. However,
it and its following work, like the work in \cite{yilmaz2004blind},
are based on a different problem from $\mathbf{x}=A\mathbf{s}$ where
other special conditions are applied, and therefore, are able to solve
for more sources than observed signals. This is also not the focus
of this paper. Other work on BSS with frequency domain analysis, like
the work in \cite{weinstein1993multi,ikeda1999method,saruwatari2001blind},
though consider the mixing relation between sources and observed signals
in frequency domain, are different from our work by assumption, model,
and algorithms.

\section{ICA Revisited\label{sec:ICArevis}}

\subsection{A Two-Step Framework\label{subsec:A-Two-Step-Framework}}

Typically, ICA consists of two steps: the ICA optimization following
a prewhitening step, where a PCA is performed. Though in most work
(\cite{comon1994independent,hyvarinen2000independent}), the prewhitening
of the input data $\mathbf{x}$ was taken for granted, it is also
well known that the purpose of the PCA is not merely to ``standardize''
$\mathbf{x}$ so as to make its covariance identity. The key is whether
to accept an additional assumption that $\mathbf{ss}^{T}=I$. This
assumption was accepted in the paper of \cite{comon1994independent}
but not in the paper of \cite{hyvarinen1999fast}. Since by the assumption
of independence, the source random variable $\mathbf{s}$ are uncorrelated.
Hence, the assumption of $\mathbf{ss}^{T}=I$ only adds an additional
condition that each source random variable has unit variance. In ICA,
the scaling of the source random variables is nondeterministic, thus
the assumption is reasonable. With this assumption, ICA becomes a
two-step algorithm, as analyzed in the following:

Expressing $A$ as its singular value decomposition (SVD) 
\begin{equation}
A=U_{A}\Sigma_{A}V_{A}^{T},\label{eq:svd1}
\end{equation}
and given that $\mathbf{s}\mathbf{s}^{T}=I$, we have 
\[
\mathbf{x}\mathbf{x}^{T}=A\mathbf{s}\mathbf{s}^{T}A^{T}=AA^{T}.
\]
i.e. 
\[
C_{X}=U_{A}\Sigma_{A}^{-2}U_{A}^{T},
\]
where $C_{X}=\mathbf{x}\mathbf{x}^{T}$. Note that $A$ is not orthogonal,
otherwise $\mathbf{x}$ are uncorrelated and no PCA is needed. Hence,
$\Sigma_{A}^{-2}$ is a diagonal matrix whose main diagonal elements
are not equal. And $C_{X}=U_{A}\Sigma_{A}^{-2}U_{A}^{T}$ is a unique
eigen decomposition. This implies that, applying PCA to $\mathbf{x}$,
we can solve for both $U_{A}$ and $\Sigma_{A}$.

Considering the SVD of the linear mixing matrix $A$, we can call
the equation 
\[
\mathbf{x}=A\mathbf{s}=U_{A}\Sigma_{A}V_{A}^{T}\mathbf{s}
\]
 a ``rotation-scaling-rotation'' procedure (up to some permutation
and reflection): $V_{A}^{T}$ is the first rotation applied to $\mathbf{s}$,
$\Sigma_{A}$ applies scalings to $\mathbf{s}$, and $U_{A}$ is the
second rotation. From above we saw that from the mathematical point
of view, the PCA step in fact solve for the second rotation $U_{A}$
and the scaling $\Sigma_{A}$. 

Therefore, a whole ICA procedure should be considered as a two-step
framework, which is also very well-known in signal processing literature
(\cite{belouchrani1998blind}): solving for the second rotation $U_{A}^{T}$
and the scaling $\Sigma_{A}$ by PCA; and then solving for the first
rotation $V_{A}^{T}$ based on other assumptions, like ``independence''
in the work in \cite{comon1994independent}, or auto-correlation matrices
being diagonal in the work in \cite{belouchrani1997blind}.

Let's call the signals after PCA as $\mathbf{z}$, i.e. 
\[
\mathbf{z}=\Sigma_{A}U_{A}^{T}\mathbf{x},
\]
and we have 
\[
\mathbf{s}=V_{A}\mathbf{z}.
\]
In the two-signal cases, the orthonormal matrix $V_{A}$ is just a
rotation matrix, up to some reflection and permutation. And in higher
dimensional cases, it is a composition of a series of rotations (and
possible reflections) within two-dimensional subspaces. This implies
an important fact, which can also be noticed from the MI-based ICA
objective functions, that:

\textit{The joint entropy of $\hat{V_{A}}\mathbf{z}$ is invariant
to rotation $\hat{V_{A}}$. }

Therefore, after the first step of an MI-based ICA, the joint entropy
is already maximized, and the second step is just a searching for
rotations that minimize the summation of each marginal entropy. This
agrees with the fact that for any MI-based ICA approach, the true
objective function is the summation of negentropies
\[
\sum_{i}J(p_{z_{i}})=\sum_{i}(H(\phi_{z_{i}})-H(p_{z_{i}}))
\]
where $p_{z_{i}}$ is the density of a random variable $z_{i}$, and
$\phi_{z_{i}}$ is the Gaussian density with the same mean and variance
as $p_{z_{i}}$. This was interpreted as ``Faraway from Gaussian
distribution implies independence'' (\cite{hyvarinen2000independent}).
Note that, during the searching of the rotation angle, the mean (standardized
as zero) and variance does not change for each $z_{i}$, so that $H(\phi_{z_{i}})$
does not change, and minimizing the negentropy is equivalent to minimizing
the sum of marginal entropies. 

This also implies that ICA only valid for the case where at most one
Gaussian component exists, since if all components are Gaussian, after
the first step, the resulted distribution is rotational symmetric,
given the assumption that $\mathbf{s}\mathbf{s}^{T}=I$. 

\begin{figure*}
\begin{centering}
\includegraphics[scale=0.1]{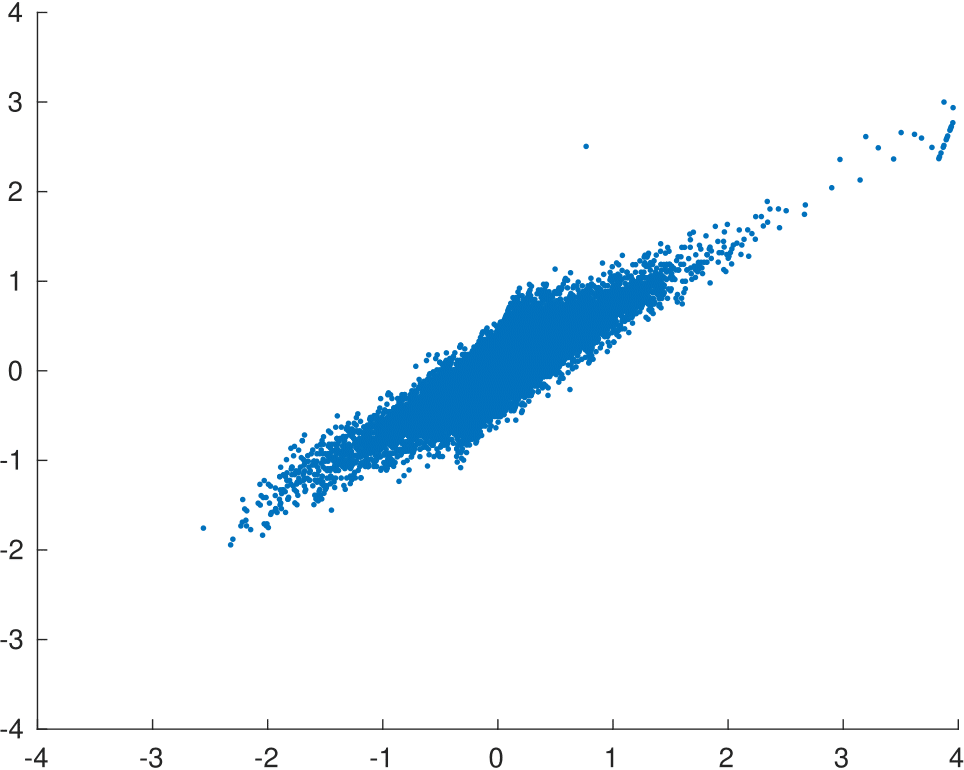}\hspace{1cm}\includegraphics[scale=0.1]{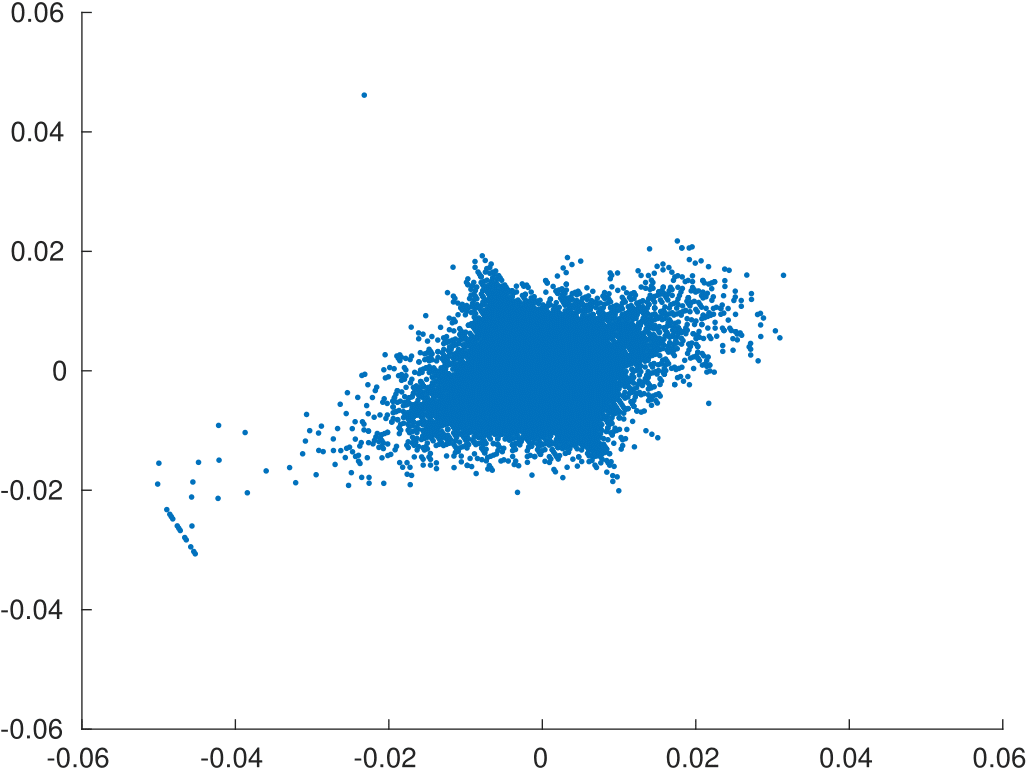}\hspace{1cm}\includegraphics[scale=0.1]{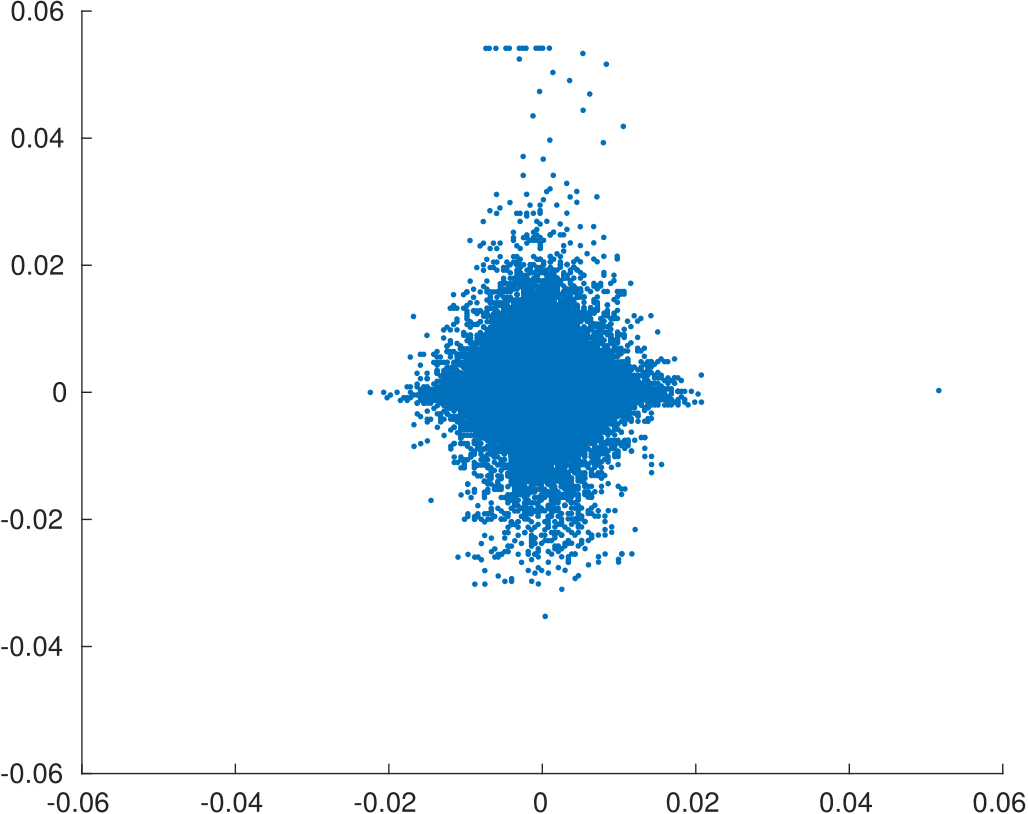}
\par\end{centering}
\caption[Scatter plots of example signals indicating the two-step framework
of ICA.]{Scatter plots of example signals indicating the two-step framework
of ICA. \textbf{Left: }the scatter plot of $\mathbf{x}(t)$; \textbf{Mid:
}the scatter plot of $\mathbf{z}(t)$; \textbf{Right: }the scatter
plot of $\mathbf{s}(t)$. From $\mathbf{x}(t)$ to $\mathbf{z}(t)$,
a rotation and a scaling was applied, while from $\mathbf{z}(t)$
to $\mathbf{s}(t)$, merely a rotation was applied.}

\label{fig:scatter}
\end{figure*}

This fact can be understood as: under the linear mixing model, uncorrelatedness
implies maximization of joint entropy, and that independence and uncorrelatedness
only differ by a series of rotations.Fig. \ref{fig:scatter} shows
an example where we can observe that the seeking of independence is
a seeking of an angle, so that each marginal distribution has as less
marginal entropies as possible.

\subsection{The Reasonability of the Independence Assumption}

In this section, we discuss the independence assumption formally.
The statistical independence can be defined from two different perspectives:
the signals being deterministic functions, or stochastic processes. 

From the stochastic process point of view, we consider the source
signals $\mathbf{s}(t)=(s_{1}(t),\dots,s_{M}(t))$ being sample functions
of continuous stationary stochastic processes $\tilde{\mathbf{s}}(t)=(\tilde{s}_{1}(t),\dots\tilde{s}_{M}(t))$.
For any positive integer $n$, pick time points $t_{1},t_{2},\dots,t_{n}\in D$
and any time interval $\Delta t\in D$, where $D$ is the time domain,
for $i=1,2,\dots,M$, the random vector 
\[
(\tilde{s_{i}}(t_{1}),\tilde{s_{i}}(t_{2}),\dots,\tilde{s_{i}}(t_{n}))
\]
 and 
\[
(\tilde{s_{i}}(t_{1}+\Delta t),\tilde{s_{i}}(t_{2}+\Delta t),\dots,\tilde{s_{i}}(t_{n}+\Delta t))
\]
 has identical distribution $p_{\tilde{s}_{i}}$. The independence
of the signals are defined as the independence of $p_{\tilde{s}_{i}}$
for $i=1,2,\dots,M$. By the ergodicity theorem of the stationary
stochastic processes, the values of the sample functions \textendash{}
the source signals \textendash{} can be used to estimate the entropy
of underlying distribution, and compute their mutual information.
Therefore, assuming independence of the distributions underlying a
set of signals is reasonable, and hence ICA can be directly applied
to BSS with the independence assumption.

\begin{figure*}
\begin{centering}
\includegraphics[scale=0.1]{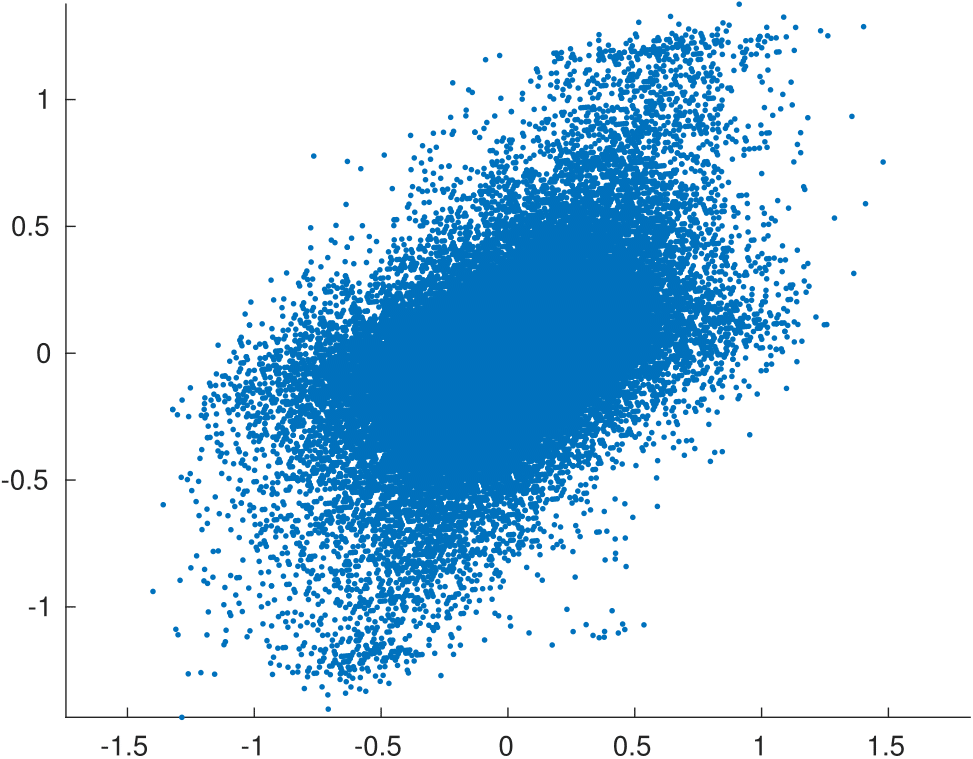}\hspace{0.3cm}\includegraphics[scale=0.1]{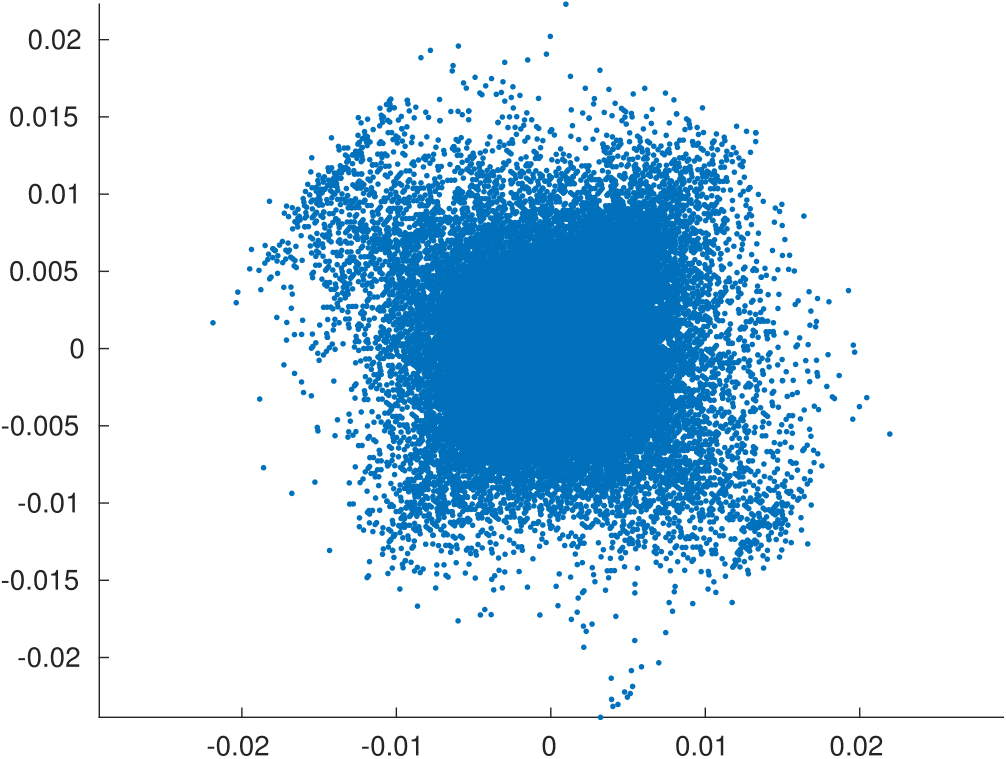}\hspace{0.3cm}\includegraphics[scale=0.1]{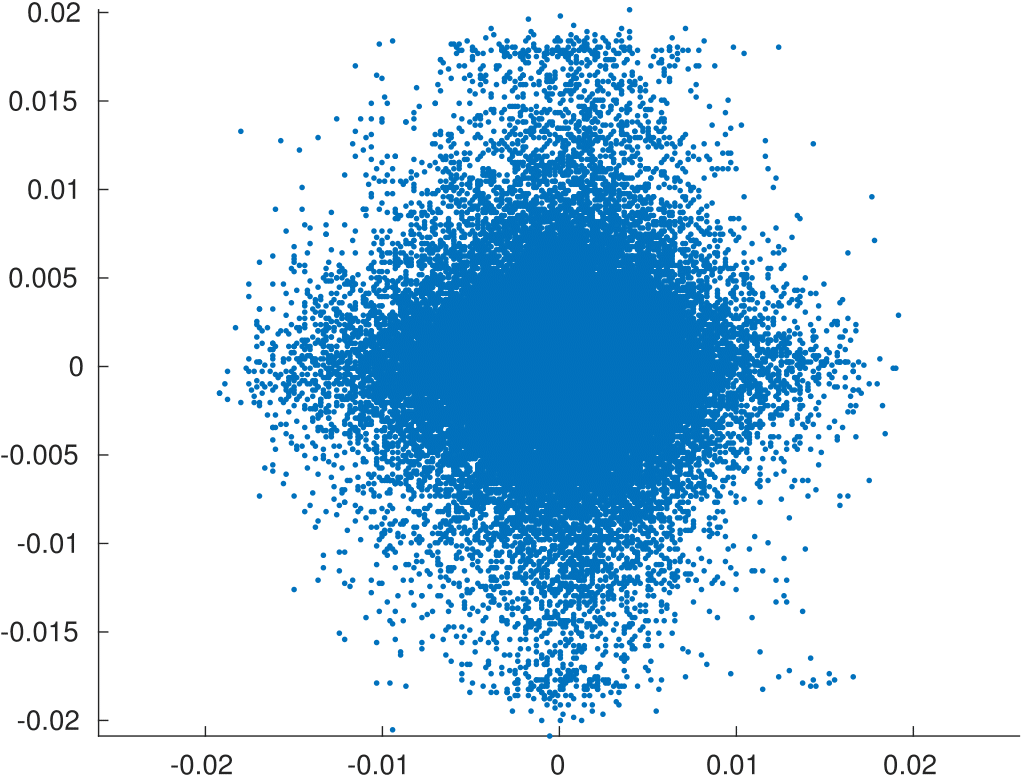}\hspace{0.3cm}\includegraphics[scale=0.1]{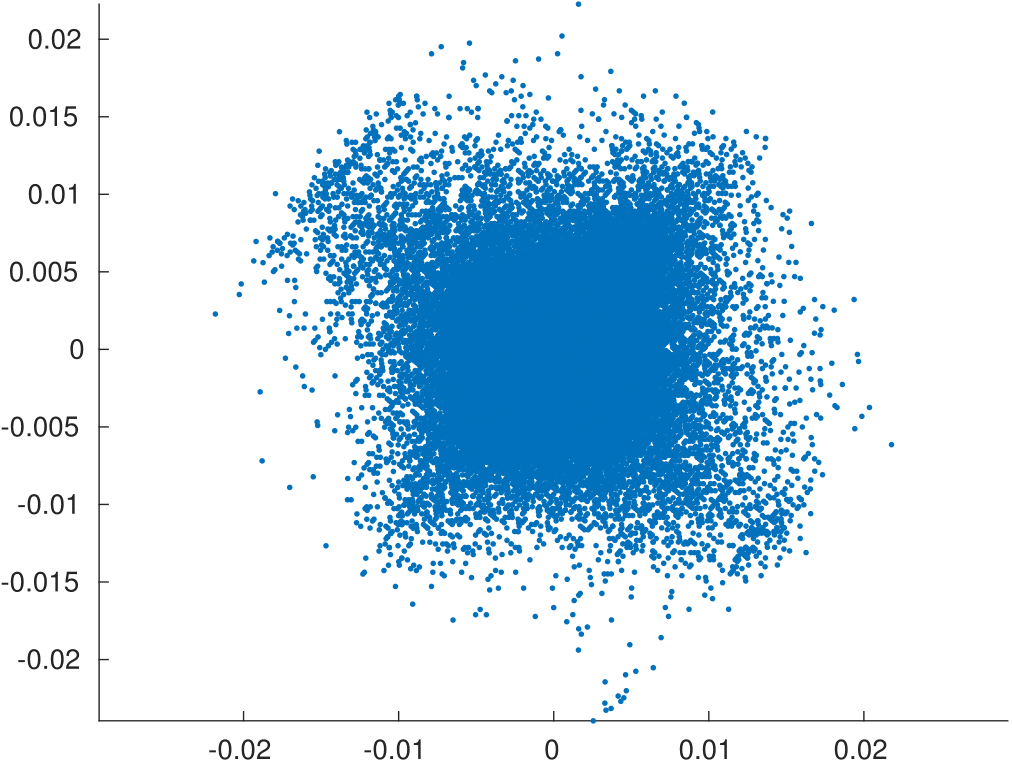}
\par\end{centering}
\caption[A Counterexample for the independence assumption.]{A Counterexample for the independence assumption. \textbf{Left: }the
scatter plot of $\mathbf{x}(t)$; \textbf{Mid-left: }the scatter plot
of $\mathbf{z}(t)$; \textbf{Mid-right: }the scatter plot of $\mathbf{s}(t)$;
\textbf{Right: }the scatter plot of $y(t)$ recovered by minimizing
sum of marginal entropies. The source signals $\mathbf{s}(t)$ has
larger sum of marginal entropies than the minimum case, and hence
MI is not able to solve for the true source signals from the mixed
observation $\mathbf{x}(t)$. }

\label{Fig. 6-3-2}
\end{figure*}

However, ``independence'' is not the truth, but just an assumption
to admit so that ICA can be applied to BSS. It is not perfect, and
has the following disadvantages: Firstly, the stochastic process model
of signals disregards the derivative information contained in the
signals. That is, if we reorder the signal sample values, there are
no difference from the stochastic process perspective. We assert that
an approach may work for more cases if it takes the derivative information
into consideration. Secondly, there exists pairs of source signals
that are generated and sampled ``independently'', but by ICA, i.e.
by the minimization of sum of marginal entropies, the original signals
may not be recovered. See Fig. \ref{Fig. 6-3-2}. This indicates the
fact that the independence assumption may not be the most reasonable
assumption for these source signals. Finally, ICA do not work for
the case where at least two source signals are Gaussian, as we mentioned
above.

\section{Applying SPC to BSS \label{sec:Applying-SPC-to}}

\subsection{The SPC Model}

Based on the effectiveness of the MI-based ICA for most cases of BSS,
as well as the disadvantages of the stochastic process model of signals
underlying ICA for BSS, we need a different approach that is based
on the deterministic function model of signals where derivatives of
the signals are available, and is closely related to MI. SPC is one
of the best choices.

The SPC model is expressed as follows:

Suppose that we have a set of 1D signals $\{s_{1}(t),s_{2}(t),\dots,s_{N}(t)\}$
defined on the domain of time $D\subset\mathbb{R}$. Consider the
mapping
\[
\mathcal{S}:D\rightarrow\mathbb{R}^{N+1}
\]
by
\[
t\mapsto(t,s_{1}(t),\dots,s_{N}(t)),
\]
and we have a 1D parametric curve $\mathcal{S}(t)$ embedded in an
$N+1$ dimensional Euclidean space. And its curve arc length is
\[
\int_{D}\sqrt{1+\sum_{i=1}^{N}(s_{i}'(t))^{2}}\mathrm{d}t.
\]

Analogizing the ISSRA objective function in 2D case, we have the following
objective function:
\begin{equation}
\mathcal{O}_{1}=\int\frac{\sqrt{N}\prod_{i=1}^{N}\sqrt[2N]{\frac{1}{N}+(s_{i}'(t))^{2}}}{\sqrt{1+\sum_{i=1}^{N}(s_{i}'(t))^{2}}}\mathrm{d}t\label{eq:o2-1}
\end{equation}
where $N$ is the number of signals. We can call it the ``Signal
Parametric Curve Relative Arc Length'', comparing to the name of
ISSRA in the ISPS model.

Note that in Eq. \ref{eq:o2-1}, comparing to ISSRA, the denominator
and numerator of the integrand are flipped. In ISSRA, the joint area
is to be minimize for getting images similar, so it is in the numerator.
And here we want to minimize $\mathcal{O}_{1}$ to get signals as
separated as possible, so we flip the integrand in order to fit this
``opposite'' problem, by putting the product of each arc length
in the numerator and the joint arc length in the denominator.

\subsection{The Relationship with MI}

In order to discuss the relationship between SPC and MI, we need to
consider the pseudo-SPC model and understand the signals being random
variables as functions of time. Consider time $t$ as a uniformly
distributed random variable, each signals $s_{i}(t)$ being a differentiable
function of $t$ is also a random variable. To estimate the joint
entropy of the ``stack of signals'', the pseudo-SPC
\[
\tilde{\mathcal{S}}:D\rightarrow\mathbb{R}^{N}
\]
by
\[
t\mapsto(s_{1}(t),\dots,s_{N}(t))
\]
is considered. The difference between SPC $\mathcal{S}$ and pseudo-SPC
$\tilde{\mathcal{S}}$ is that the first dimension $t$ does not appear
in $\tilde{\mathcal{S}}$, and $\tilde{\mathcal{S}}$ is not injective,
similar to the relationship between ISPS and pseudo-ISPS.

Unfortunately, similar to the fact that MI-based registration approach
is not applied to groupwise case, because of the disagreement of dimensionality,
the Lebesgue measure of $\tilde{\mathcal{S}}(t)$ embedded in $\mathbb{R}^{N}$
is zero, and the joint density does not exist. And from the statistics
point of view, it is also clear that, since each signal is a function
of $t$, there is no independence defined for the set of all signals.
This implies that under the pseudo-SPC point of view, MI is not able
to be computed to solve the BSS problem.

However, in Section \ref{subsec:A-Two-Step-Framework}, we discuss
the two-step framework of the ICA problem. We pointed out that in
ICA the joint density and joint entropy is never considered. In the
second step, no joint entropy is computed, but just the sum of marginal
entropies. Fortunately, in the pseudo-SPC perspective, the 1D marginal
entropy of each signal is well-defined. And by the close relationship
between it and the SPC model, we are still able to apply $\mathcal{O}_{1}$
to BSS to approximate the ``MI'', i.e. the sum of marginal entropies,
to solve for unmixed signals.

In fact, looking at $\mathcal{O}_{1}$ carefully, we notice that the
joint arc length (the denominator) is also invariant to rotations,
which means that in the second step of ICA where different rotation
matrices are applied, the denominator does not change either. This
also meets the fact that $\mathcal{O}_{1}$ is closely related to
MI, where the joint entropy part does not change with respect to rotations.
And we can simplify $\mathcal{O}_{1}$ to get 
\[
\mathcal{O}_{2}=\int\prod_{i=1}^{N}\sqrt{1+(s_{i}'(t))^{2}}\mathrm{d}t
\]
where only the marginal arc lengths are computed. Clearly, $\mathcal{O}_{2}$
is related to the true objective function, the sum of marginal entropies,
in the traditional MI approaches for ICA, and can be considered as
the objective function derived from the SPC model.

\subsection{Geometrical Objective Functions for BSS\label{subsec:Geometrical-Objective-functions}}

Applying SPC to BSS, and considering the two-step framework of ICA,
we proposed the objective function $\mathcal{O}_{2}$, the product
of marginal arc lengths, by its close relationship with the objective
function of traditional MI approach. Hence, in the second step of
ICA, given $\mathbf{z}(t)$ as the inputs, we can apply a rotation
matrix $R$ to get 
\[
\mathbf{y}(t)=R\mathbf{z}(t)
\]
and computes the objective functions of $\mathbf{y}(t)$ to solve
for best $\hat{\mathbf{y}}(t)$ that approximates $\mathbf{s}(t)$
best. The optimization can be done either by brute-force search, or
gradient descent algorithm since the objective function $\mathcal{O}_{2}$
is smooth and convex (see Section \ref{sec:6-Experiments}). In this
paper, for simplicity we only do brute-force search for each objective
function for comparison.

We also propose some other objective functions which have similar
structures as $\mathcal{O}_{2}$:

\[
\mathcal{O}_{3}=\int\prod_{i=1}^{N}|y_{i}'(t)|\mathrm{d}t
\]

\[
\mathcal{O}_{4}=\int\log\prod_{i=1}^{N}|y_{i}'(t)|\mathrm{d}t
\]
\[
\mathcal{O}_{5}=\sum_{i=1}^{N}\int\sqrt{1+(y_{i}'(t))^{2}}\mathrm{d}t
\]

All these above objective functions come from the arc lengths of each
signals, and are named geometrical objective functions for BSS.

Comparing with the objective function of sum of marginal entropies,
the advantages of these functions are: they computes easier and faster
than estimation of densities; they consider the derivative information
of signals; they do not assume the independence, and work for the
case where sources are not independent (for example, the counterexample
shown in Fig. \ref{Fig. 6-3-2}).

Other than this dissertation, there do exist previous work that proposed
other functions approximating the traditional MI objective functions.
The most famous ones are the following, proposed in the work in \cite{hyvarinen1999fast}:

\[
G_{1}(y)=\frac{1}{a_{1}}\log\cosh(a_{1}y)
\]
\[
G_{2}(y)=-\frac{1}{a_{2}}\exp(-a_{2}y^{2}/2)
\]
\[
G_{3}(y)=\frac{1}{4}y^{4}
\]
where $a_{1}$ and $a_{2}$ are hyperparameters.

In Section \ref{subsec:The-Behaviors-of} we show the function graph
of each of the above objective functions. The results showed that
all these geometrical objectives and the contrast functions agree
at similar global minimum, up to some approximation error, which indicates
that all these objective functions have similar behaviors in the BSS
problems, and are effective approaches. However, among them, the geometrical
objectives have significant better precision, especially $\mathcal{O}_{3}$
and $\mathcal{O}_{5}$, which indicates that the geometrical objective
functions not only share good properties with the contrast functions,
but also have better performance. Therefore, they are competitive
replacements of contrast functions, and have both theoretical and
practical potentials.

\section{Frequency Domain Approaches and the New Assumption\label{sec:FT-PCA}}

\subsection{Motivation\label{subsec:Motivation}}

Among the new objective functions proposed above, $\mathcal{O}_{3}(R)=\int|y'_{1}(t)y'_{2}(t)|\mathrm{d}t$
has the simplest formula. An immediate question then comes up: does
it work if we simplify it further by taking away the absolute value
sign, i.e. $\tilde{\mathcal{O}}_{3}(R)=\int y'_{1}(t)y'_{2}(t)\mathrm{dt}$?
From the experiment results in Section \ref{subsec:The-Behaviors-of},
we can observe that it has worse performance than $\mathcal{O}_{3}$,
but its error was acceptable for a practical BSS task.

For different $s_{1}(t)$ and $s_{2}(t)$, most likely $\int(s'_{1}(t))^{2}\mathrm{d}t\neq\int(s'_{2}(t))^{2}\mathrm{d}t$,
then suppose that the minimization of $\tilde{\mathcal{O}}_{3}(R)$
leads to 
\[
\min\int y'_{1}(t)y'_{2}(t)\mathrm{d}t=\int s'_{1}(t)s'_{2}(t)\mathrm{d}t=0,
\]
This induces the actual assumption of $\tilde{\mathcal{O}_{3}}$,
other than approximating MI. Accepting this assumption, we can solve
BSS by solving a PCA problem on the derivatives of given signals $z_{1}(t)$
and $z_{2}(t)$:

Suppose that $\hat{R}$ is the correct rotation matrix to be solved,
i.e. $\mathbf{s}(t)=\hat{R}\mathbf{z}(t)$. Taking derivatives on
both sides, we have
\begin{equation}
\mathbf{s}'(t)=\hat{R}\mathbf{z}'(t).\label{eq:4}
\end{equation}
Hence, 
\[
\int\mathbf{z}'(t)(\mathbf{z}'(t))^{T}\mathrm{d}t=\hat{R}^{T}\int\mathbf{s}'(t)(\mathbf{s}'(t))^{T}\mathrm{d}t\hat{R}.
\]
Since $\int(s'_{1}(t))^{2}\mathrm{d}t\neq\int(s'_{2}(t))^{2}\mathrm{d}t$
and $\int s'_{1}(t)s'_{2}(t)\mathrm{d}t=0$, $\int\mathbf{s}'(t)(\mathbf{s}'(t))^{T}\mathrm{d}t$
is a nontrivial diagonal matrix. Therefore, similar to the first PCA
step, by eigen decomposition of $\int\mathbf{z}'(t)(\mathbf{z}'(t))^{T}\mathrm{d}t$,
we are able to get $\hat{R}$. We call this approach Derivative-PCA.

\begin{figure*}
\begin{centering}
\includegraphics[scale=0.1]{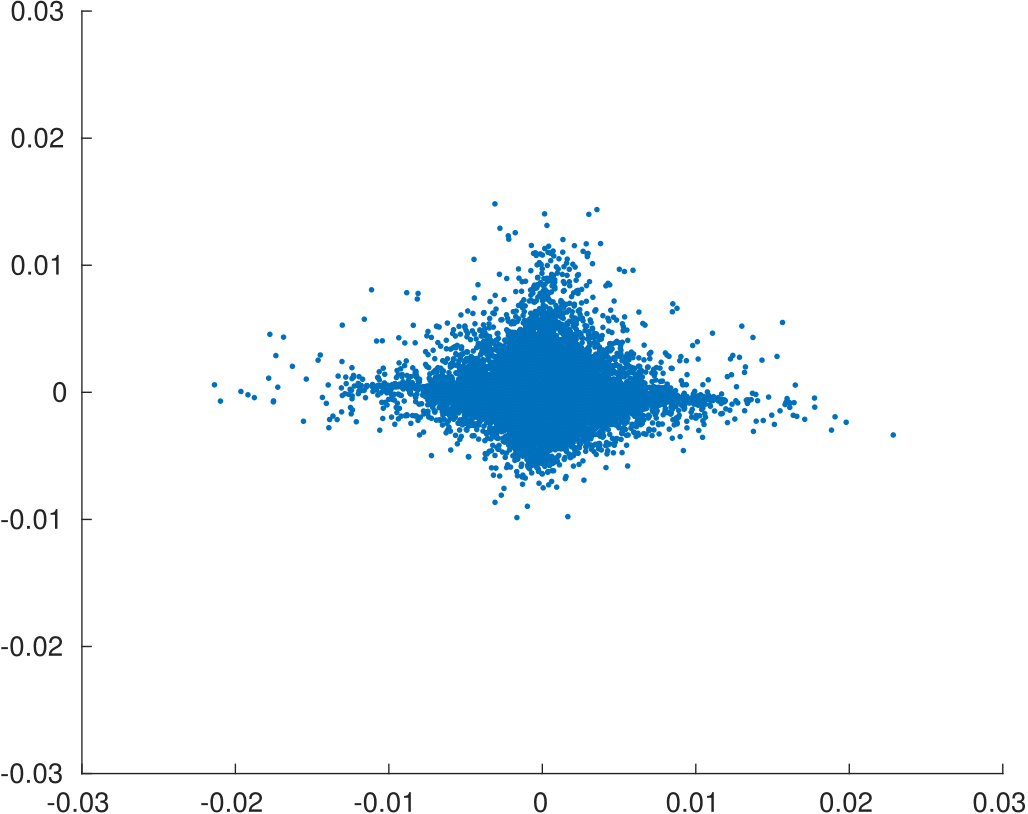}\hspace{1cm}\includegraphics[scale=0.1]{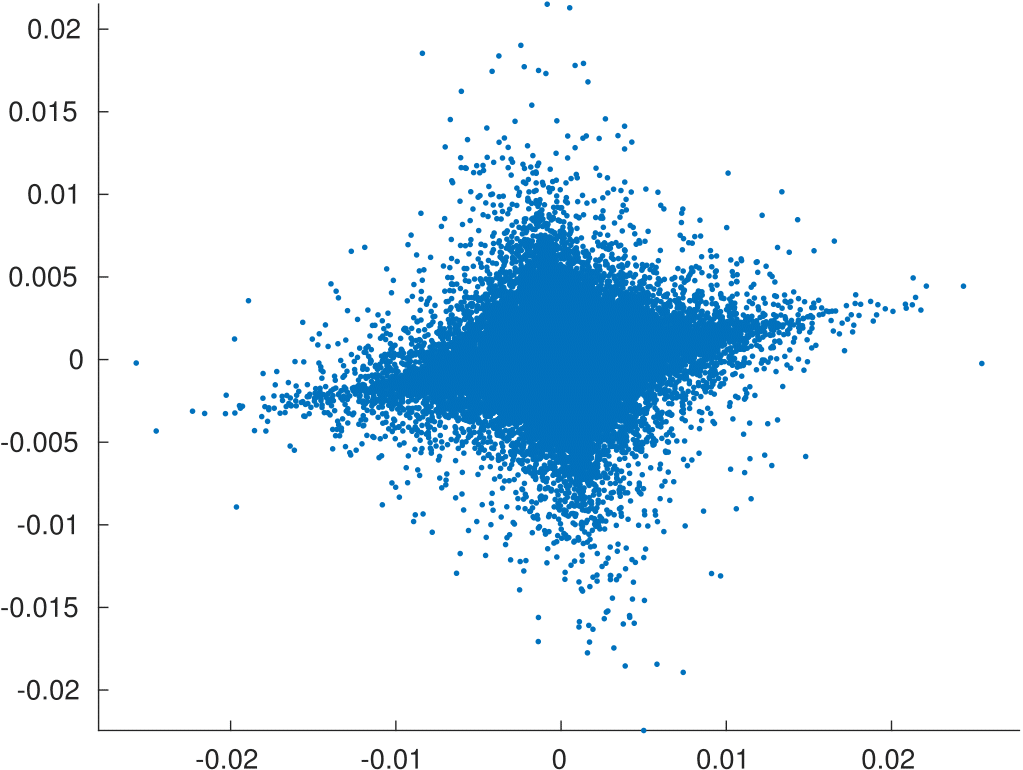}\hspace{1cm}\includegraphics[scale=0.1]{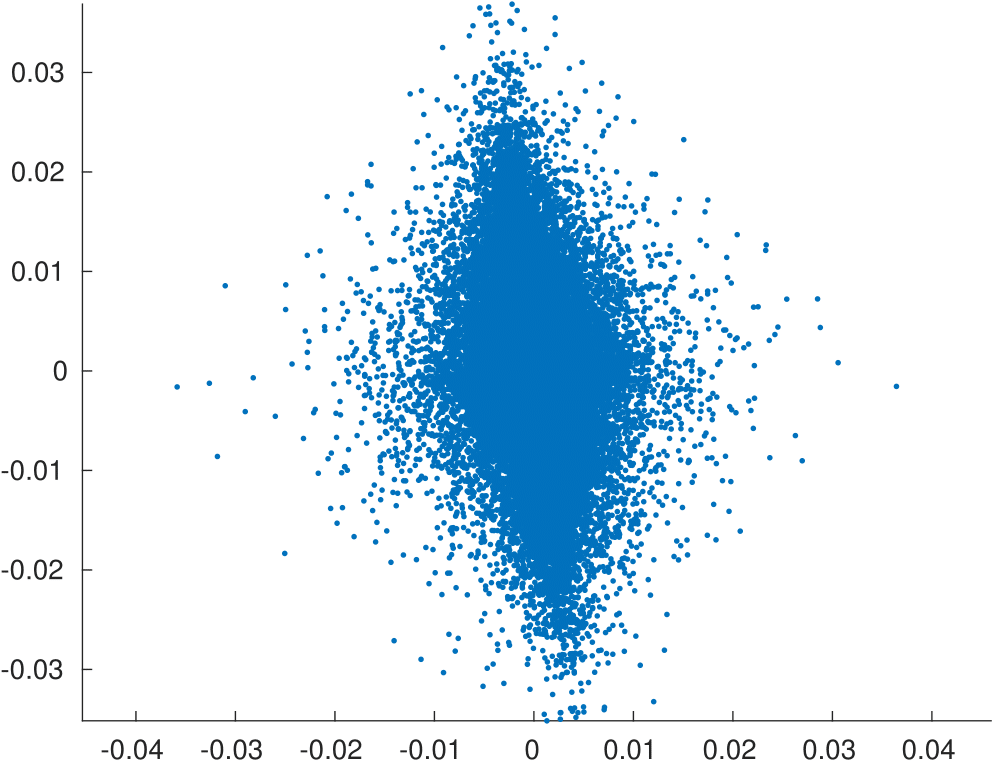}
\par\end{centering}
\caption[Counterexamples of the assumption of derivative orthogonality.]{\textbf{ }Counterexamples of the assumption of derivative orthogonality.
Each figure shows a scatter plot of the derivatives of a pair of source
signals. From the figures we can observe that the derivatives are
not orthogonal. The covariance of each pair of derivatives are: (in
order) 0.0195, 0.0881, 0.0580, which are relatively large.}

\label{fig.ind-ce}
\end{figure*}

The Derivative-PCA approach is based on the assumption that $\int(s'_{1}(t))^{2}\mathrm{d}t\neq\int(s'_{2}(t))^{2}\mathrm{d}t$
and $\int s'_{1}(t)s'_{2}(t)\mathrm{d}t=0$, $\int\mathbf{s}'(t)(\mathbf{s}'(t))^{T}\mathrm{d}t$.
In other words, for a BSS problem, suppose that we do not have any
assumption on independence, but an assumption on orthogonality of
derivatives of the source signals, and then we are able to solve for
the second rotation by another PCA step on the derivative signals.
As what we discuss in Section \ref{sec:Introduction}, a BSS problem
is highly open, and it can only be solved with restrictions based
on assumptions. However, assumptions are not true or false. An assumption
works if it meets the real cases. Just like what we discussed above:
in most cases the independence assumption works, but there are also
counterexamples. Similarly, if most source signals that are sampled
``independently'' have orthogonal derivatives, then the above approach
would give correct approximations to the sources. Unfortunately, practically
speaking, it is easier to find counterexamples for the assumption
of orthogonal derivatives than the assumption of independence. Fig.
\ref{fig.ind-ce} shows one of the examples where the original signals
do not have orthogonal derivatives. Empirically, we may assert that
the orthogonality on derivative signals is not good enough as an assumption.

However, this approach is inspiring, which gives the motivation of
this section: finding a reasonable assumption that can construct a
PCA-like structure for solving BSS, bypassing the optimization procedure.
This approach falls in the second-order-statistics category for BSS.
And as we discussed in Section \ref{sec:Previous-Work}, AMUSE, SOBI,
and STFD are well-known approaches in this category. In the following,
we propose a new approach FT-PCA following this idea, and discuss
the reasonability of its assumption comparing with AMUSE and SOBI.

\subsection{The Fourier Transform Approach\label{subsec:The-Fourier-Transform}}

The goal of the second step is to solve for $\mathbf{s}(t)$ and $\hat{R}$
from $\mathbf{s}(t)=\hat{R}\mathbf{z}(t)$. Applying Fourier transform
(FT) on both sides, we have
\[
\mathbf{S}(\omega)=\hat{R}\mathbf{Z}(\omega),
\]
where $\omega\in\mathbb{R}$ is the frequency, $\mathbf{S}(\omega)$
is the FT of $\mathbf{s}(t)$, and $\mathbf{Z}(\omega)$ is the FT
of $\mathbf{z}(t)$. 

By Parseval's Theorem, we know that
\[
\int|Z_{i}(\omega)|^{2}\mathrm{d}\omega=\int(z_{i}(t))^{2}\mathrm{d}t=1
\]
\[
\int|S_{i}(\omega)|^{2}\mathrm{d}\omega=\int(s_{i}(t))^{2}\mathrm{d}t=1
\]
for $t=1,2$, and
\[
\int Z_{1}(\omega)\overline{Z_{2}(\omega)}\mathrm{d}\omega=\int z_{1}(t)z_{2}(t)\mathrm{d}t=0
\]
\[
\int S_{1}(\omega)\overline{S_{2}(\omega)}\mathrm{d}\omega=\int s_{1}(t)s_{2}(t)\mathrm{d}t=0.
\]
 Therefore, $\int\mathbf{Z}(\omega)(\mathbf{Z}(\omega))^{H}\mathrm{d}\omega=\int\mathbf{S}(\omega)(\mathbf{S}(\omega))^{H}\mathrm{d}\omega=\mathbf{I}$
are both the identity matrix.

The above transformation gives trivial results since $\hat{R}$ is
not able to be solved from 
\[
\int\mathbf{Z}(\omega)(\mathbf{Z}(\omega))^{H}\mathrm{d}\omega=\hat{R}^{T}\int\mathbf{S}(\omega)(\mathbf{S}(\omega))^{H}\mathrm{d}\omega\hat{R}
\]
 which is equivalent to $\mathbf{I}=\hat{R}^{T}\mathbf{I}\hat{R}=\mathbf{I}$.
In fact, any change of basis applied to the function space of the
signals have similar results, due to the Parseval's Theorem. However,
inspired by the above derivative orthogonality assumption, we can
apply kernel tricks as follows:

Multiplying both sides of $\mathbf{S}(\omega)=\hat{R}\mathbf{Z}(\omega)$
by a certain complex function $\phi(\omega)$ that is nonzero on a
set with positive Lebesgue measure, we have
\[
\phi(\omega)\mathbf{S}(\omega)=\phi(\omega)\hat{R}\mathbf{Z}(\omega).
\]
Let $K(\omega)=\phi(\omega)\overline{\phi(\omega)}$ be the kernel
function, then the elements of the covariance matrices $\int K(\omega)\mathbf{Z}(\omega)(\mathbf{Z}(\omega))^{H}\mathrm{d}\omega$
and $\int K(\omega)\mathbf{S}(\omega)(\mathbf{S}(\omega))^{H}\mathrm{d}\omega$
become inner products of FT of signals in the kernel space defined
by $K(\omega)$, and we have\textit{
\[
\int K(\omega)\mathbf{Z}(\omega)(\mathbf{Z}(\omega))^{H}\mathrm{d}\omega=\hat{R}^{T}\int K(\omega)\mathbf{S}(\omega)(\mathbf{S}(\omega))^{H}\mathrm{d}\omega\hat{R}.
\]
}For convenience, let us name each of the elements in the above matrices
as follows:

\[
\mathbf{S}_{ij}=\int K(\omega)S_{i}(\omega)\overline{S_{j}(\omega)}\mathrm{d}\omega
\]
and
\[
\mathbf{Z}_{ij}=\int K(\omega)Z_{i}(\omega)\overline{Z_{j}(\omega)}\mathrm{d}\omega
\]
under the case where $K(\omega)$ has no ambiguity, then we can write
that
\begin{equation}
\int K(\omega)\mathbf{Z}(\omega)(\mathbf{Z}(\omega))^{H}\mathrm{d}\omega=\left(\begin{array}{cc}
\mathbf{Z}_{11} & \mathbf{Z}_{12}\\
\mathbf{Z}_{21} & \mathbf{Z}_{22}
\end{array}\right)\label{eq:5}
\end{equation}
and 
\[
\int K(\omega)\mathbf{S}(\omega)(\mathbf{S}(\omega))^{H}\mathrm{d}\omega=\left(\begin{array}{cc}
\mathbf{S}_{11} & \mathbf{S}_{12}\\
\mathbf{S}_{21} & \mathbf{S}_{22}
\end{array}\right).
\]

Suppose there exists a kernel space defined by $K(\omega)$, so that $\mathbf{S}_{11}\neq\mathbf{S}_{22}$ 
and $\mathbf{S}_{12}=0$, then $\int K(\omega)\mathbf{S}(\omega)(\mathbf{S}(\omega))^{H}\mathrm{d}\omega$ 
is a nontrivial diagonal matrix, and $\hat{R}$ can be solved by eigen
decomposition of $\int K(\omega)\mathbf{Z}(\omega)(\mathbf{Z}(\omega))^{H}\mathrm{d}\omega$
by the uniqueness property of eigen decompositions. Formally, suppose
that $\int K(\omega)\mathbf{S}(\omega)(\mathbf{S}(\omega))^{H}\mathrm{d}\omega$
is a diagonal but not the identity matrix, the eigen decomposition
of $\int K(\omega)\mathbf{Z}(\omega)(\mathbf{Z}(\omega))^{H}\mathrm{d}\omega$
can be written as 
\[
\int K(\omega)\mathbf{Z}(\omega)(\mathbf{Z}(\omega))^{H}\mathrm{d}\omega=E^{T}\Lambda E.
\]
Then $\int K(\omega)\mathbf{S}(\omega)(\mathbf{S}(\omega))^{H}\mathrm{d}\omega$
and $\Lambda$ only differ by row switching, and $E$ and $\hat{R}$
only differ by row switching and signs. This approach of solving for
the second rotation in BSS is called FT-PCA.

Note that in the ideal case where $\mathbf{S}_{12}=\mathbf{S}_{21}=0$,
$\int K(\omega)\mathbf{S}(\omega)(\mathbf{S}(\omega))^{H}\mathrm{d}\omega$
is a real matrix. And since $\hat{R}$ is real, $\int K(\omega)\mathbf{Z}(\omega)(\mathbf{Z}(\omega))^{H}\mathrm{d}\omega$
is also a real matrix. Therefore, under the ideal kernel $K(\omega)$,
we only need to consider the real part of the matrix $\int K(\omega)\mathbf{Z}(\omega)(\mathbf{Z}(\omega))^{H}\mathrm{d}\omega$,
and consider the imaginary part as error.

The key points of FT-PCA are the reasonability of assuming the kernel
orthogonality, i.e. $\mathbf{S}_{ij}=\int K(\omega)S_{i}(\omega)\overline{S_{j}(\omega)}\mathrm{d}\omega=0$
for $i\neq j$, and if this is reasonable, how to find the kernel
$K(\omega)$.

From Section \ref{subsec:Motivation} we know that the approach of
Derivative-PCA works for some inputs, but does not work for others.
Note that by applying FT to both sides of Eq. \ref{eq:4}, we have
\[
\omega\mathbf{S}(\omega)=\omega\hat{R}\mathbf{Z}(\omega),
\]
and the Derivative-PCA approach is just a special case of FT-PCA where
the kernel $K_{1}(\omega)=\omega^{2}$. This candidate kernel works
for some inputs, but not perfect since there exist counterexamples.

\begin{figure}
\begin{centering}
\includegraphics[scale=0.38]{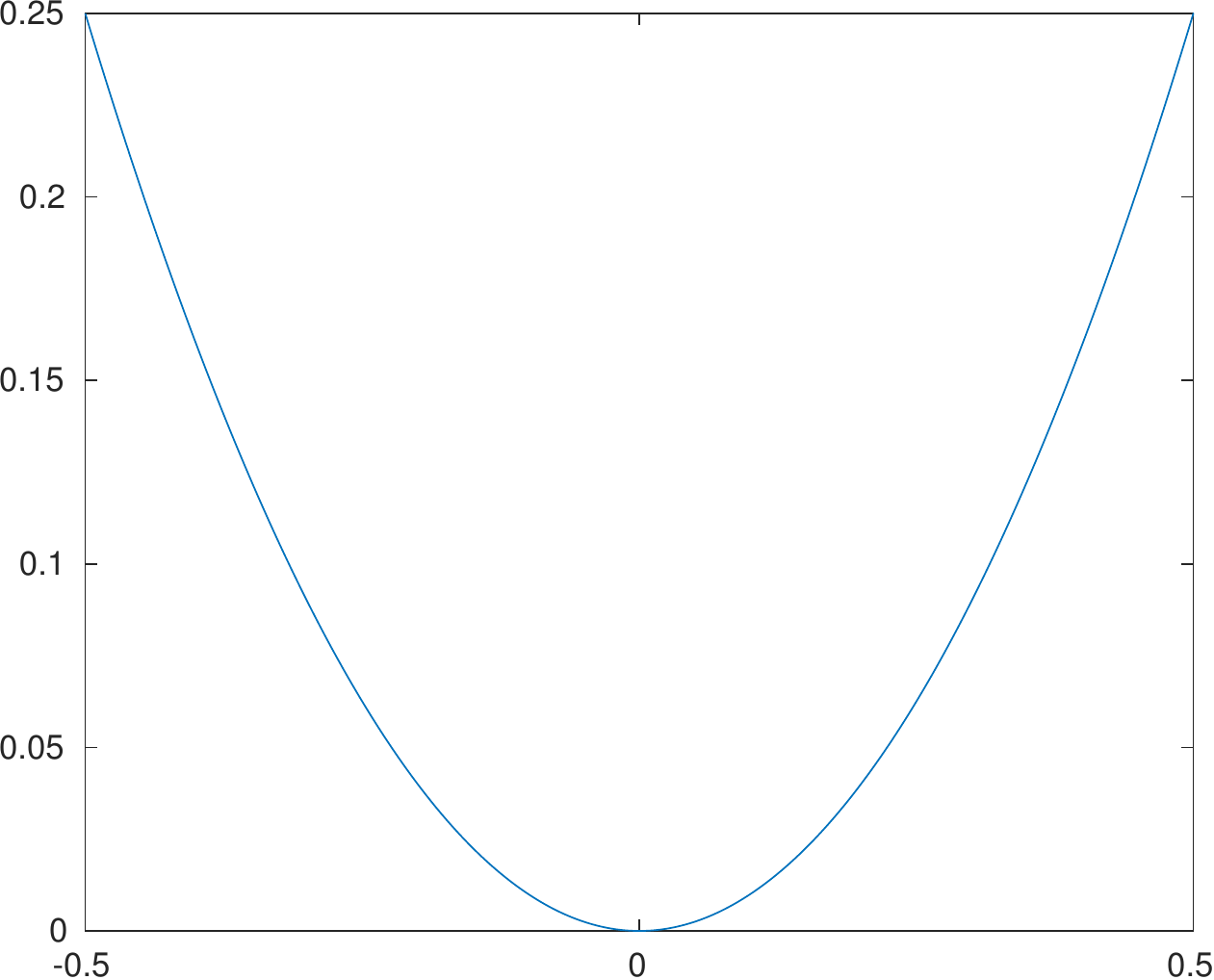}\hspace{2cm}\includegraphics[scale=0.4]{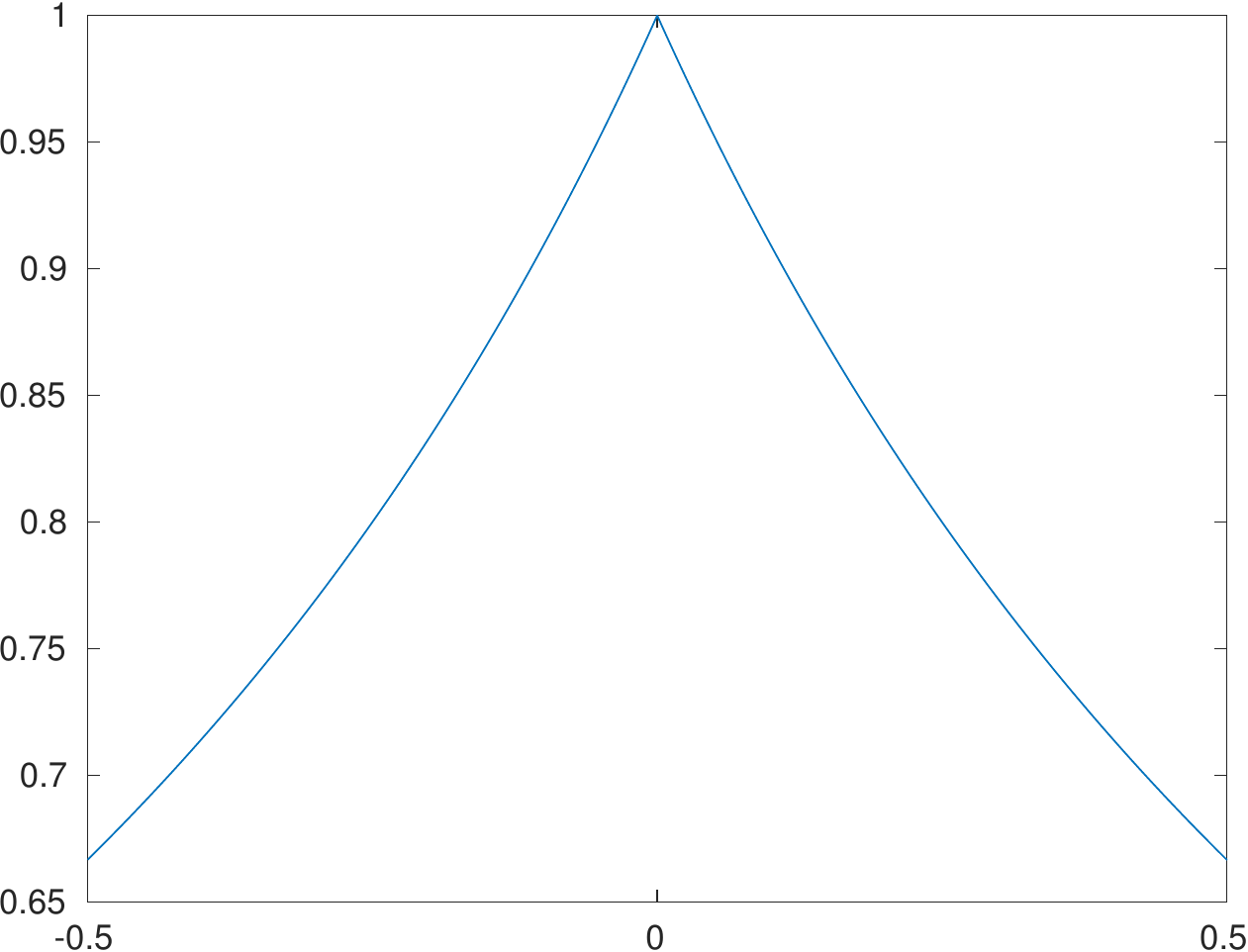}
\par\end{centering}
\caption[The function graph of $K_{1}(\omega)$ and $K_{2}(\omega)$.]{ The function graph of $K_{1}(\omega)$ and $K_{2}(\omega)$.\textbf{
Left: }the function graph of $K_{1}(\omega)$ which gives higher weights
to high frequency parts; \textbf{Right: }the function graph of $K_{2}(\omega)$
which gives higher weights to lower frequency parts. By extending
$K_{2}(\omega)$ to $K_{3}(\omega)$, the center shift $\omega_{0}$
grants the kernel focusing on frequency parts defined by user.}

\label{fig.kernelshape}
\end{figure}

By noticing the function graph of $K(\omega)=\omega^{2}$, we observe
that this kernel is similar to a window function that focuses on the
high frequency intervals of the source signals, and hence, an immediate
alternative option comes up:
\[
K_{2}(\omega)=\frac{1}{1+|\omega|}
\]
which grants low frequency parts of the signals more weights. See
Fig. \ref{fig.kernelshape}. Certainly, we can generalize it by
\begin{equation}
K_{3}(\omega)=\frac{1}{1+|\omega-\omega_{0}|}\label{eq:kernel}
\end{equation}
where $\omega_{0}$ is the center of this window-like function. By
picking different $\omega_{0}$'s, $K(\omega)$ focuses on different
intervals of the frequency domain by giving that interval higher weights,
so as to grant $\mathbf{S}_{12}$ close to zero and $\mathbf{S}_{11}\neq\mathbf{S}_{22}$.
If there exists an ideal $\omega_{0}$ so that the kernel orthogonality
assumption holds, then FT-PCA can theoretically solve the BSS problems.
In Section \ref{subsec:6-The-Reasonability-of} we show that the kernel
orthogonality assumption is reasonable, and in Section \ref{subsec:The-Heuristic-Strategy}
we show that the ideal $\omega_{0}$ is not available, but provide
a strategy to search for good $\omega_{0}$ to practically solve BSS
using FT-PCA.

With $\omega_{0}$ as the hyper-parameter, we have the FT-PCA Algorithm
described as \ref{algo}:

\begin{algorithm}
\textbf{Input: }signals $\mathbf{x}(t)$, the hyper parameter $\omega_{0}$.

\textbf{First Step:}

1. Centering $\mathbf{x}$ by $\mathbf{x}\leftarrow\mathbf{x}-\bar{\mathbf{x}}$;

2. Let $\mathbf{x}=U\Sigma V^{\ensuremath{T}}$be the SVD of $\mathbf{x}$;

3. Compute $\mathbf{z}=\Sigma^{-1}U^{T}\mathbf{x}$;

\textbf{Second Step:}

4. Compute the FT of $\mathbf{z}$ as $\mathbf{Z}(\omega)$;

5. $K(\omega)=\frac{1}{1+|\omega-\hat{\omega_{0}}|}$;

6. Compute eigen decomposition of the matrix $\mathrm{Re}(\int K(\omega)\mathbf{Z}(\omega)(\mathbf{Z}(\omega))^{H}\mathrm{d}\omega)=E\Lambda E^{T}$;

7. $\mathbf{S}(\omega)=E^{T}\mathbf{Z}(\omega)$;

\smallskip{}

\begin{raggedright}
8. Compute the inverse FT of $\mathbf{S}(\omega)$ as $\mathbf{s}$.
\par\end{raggedright}
\textbf{Output: }the separated signals $\mathbf{s}(t)$.

\caption{The FT-PCA algorithm with $\omega_{0}$ as a hyper-parameter.}

\label{algo}
\end{algorithm}

\subsection{The Reasonability of the Assumption\label{subsec:6-The-Reasonability-of}}

Firstly, let's discuss the assumptions of the previous approaches.
The assumptions of the second-order-statistics approaches SOBI and
AMUSE are deficient. The assumption of AMUSE is that
\[
\lim_{T\rightarrow\infty}\frac{1}{T}\sum_{t=1}^{T}\mathbf{s}(t+\tau)\mathbf{s}(t)^{*}=\mathrm{diag}[\rho_{1}(\tau),\dots,\rho_{n}(\tau)]
\]
for a certain $\tau$. And the assumption of SOBI shares the same
formula, except that it is true for a set of different $\tau$'s.
Both algorithms did not give a clear approach to determine which $\tau$
satisfies the assumption. If we assume that for any $\tau$, $s_{i}(t+\tau)s_{j}(t)=0$,
it implies that the correlation of the two functions
\[
s_{i}(t)\star s_{j}(t)=\int\overline{f(\tau)}g(t+\tau)\mathrm{d}\tau=0
\]
and hence,
\[
\overline{S_{i}(\omega)}S_{j}(\omega)=0
\]
where $S_{i}(\omega)$ is the Fourier transform of $s_{i}(t)$. Obviously,
except for special cases, the multiplication of the Fourier transform
of two signals cannot be a zero function. Therefore, this assumption
is too strong. If we cannot assume the autocorrelation being zero
for any time shift $\tau$, the assumption is not complete since we
are not able to aware of how to select the correct time shifts. And
most importantly, it does not make physical sense why such time shifts
should exist so that the autocorrelation of the source signals are
zero. The SOBI algorithm introduces the joint diagonalization strategy,
to select a collection of $\tau$'s, and compute based on the average
pattern of the covariance matrix, in order to bypass the deficiency
of their assumption. The case of STFD is similar, where the assumption
is that special time and frequency shifts $t$ and $f$ can be selected
so that the covariance matrix has the diagonal structure. This is
not guaranteed by theory, since only very strong assumption can guarantee
the diagonal structure for any shifts $t$ and $f$. And the algorithm
has to apply joint diagonalization. Therefore, since the assumption
of SOBI and STFD are either too strong or not applicable, their algorithms
are heuristic.

Comparing to the incompleteness of the assumptions of previous approaches,
we discuss the assumption of FT-PCA in the following.

The first issue that we need to discuss is the existence of $K(\omega)$.
Suppose that $|S_{1}(\omega)|^{2}\neq|S_{2}(\omega)|^{2}$ on a subset
$D\subset\mathbb{R}$ where $\mathrm{m}(D)>0$. (Without loss of generality,
we can suppose that $|S_{1}(\omega)|^{2}-|S_{2}(\omega)|^{2}>0$ on
$D$, since if there exists a subset with positive Lebesgue measure
such that $|S_{1}(\omega)|^{2}\neq|S_{2}(\omega)|^{2},$we can always
pick a subset of it such that $|S_{1}(\omega)|^{2}>|S_{2}(\omega)|^{2}$
or $|S_{1}(\omega)|^{2}<|S_{2}(\omega)|^{2}$.) Then the nonnegative
kernel function $K(\omega)$ with $\int K(\omega)\mathrm{d}\omega>0$
exists so that $\mathbf{S}_{11}\neq\mathbf{S}_{22}$, since we can
always pick 
\[
K(\omega)=1_{D}(\omega)
\]
 where 
\[
\int K(\omega)\mathrm{d}\omega=\mathrm{m}(D)>0.
\]

On the other hand, suppose that $|S_{1}(\omega)|^{2}\neq|S_{2}(\omega)|^{2}$
only on a subset of the frequency domain with zero Lebesgue measure,
then $K(\omega)$ does not exist. Because 
\[
\int(|S_{1}(\omega)|^{2}-|S_{2}(\omega)|^{2})^{2}\mathrm{d}\omega=0,
\]
and hence, for any $K(\omega)$, \textit{
\begin{eqnarray*}
 &  & |\mathbf{S}_{11}-\mathbf{S}_{22}|\\
 & = & |\int K(\omega)(|S_{1}(\omega)|^{2}-|S_{2}(\omega)|^{2})\mathrm{d}\omega|\\
 & \le & \int|K(\omega)(|S_{1}(\omega)|^{2}-|S_{2}(\omega)|^{2})|\mathrm{d}\omega\\
 & \le & \bigg(\int(K(\omega))^{2}\mathrm{d}\omega\bigg)^{1/2}\bigg(\int(|S_{1}(\omega)|^{2}-|S_{2}(\omega)|^{2})^{2}\mathrm{d}\omega\bigg)^{1/2}\\
 & = & 0
\end{eqnarray*}
}This indicates that, if the two source signals have the same energy
density almost everywhere, no kernel functions exist so that the two
signals can be separated by FT-PCA. Therefore, we have an necessary
condition for the source signals: FT-PCA does not work for signals
whose power spectral densities are the same. This necessary condition
excludes the cases where the source signals are too close, for example
$s_{1}(t)=\sin t$ and $s_{2}(t)=\cos t$. 

Secondly, suppose that there exists an interval $D\subset\mathbb{R}$
such that all the following conditions are satisfied:
\begin{enumerate}
\item $S_{1}(\omega)\neq0$ on a subset $D_{1}\subset D$ with $m(D_{1})>0$ 
\item $S_{2}(\omega)\neq0$ on a subset $D_{2}\subset D$ with $m(D_{2})>0$ 
\item $D_{1}\cap D_{2}=\emptyset$
\end{enumerate}
Then we immediately have that $\int_{D}|S_{1}(\omega)|^{2}\mathrm{d}\omega\neq\int_{D}|S_{1}(\omega)|^{2}\mathrm{d}\omega$
and $\int_{D}S_{1}(\omega)S_{2}(\omega)\mathrm{d}\omega=0$. The physical
meaning of the these conditions can be interpreted directly: if there
exists an interval on which the two source signals have exclusive
spectral density, then FT-PCA works. This sufficient condition gives
us a clear intuition of the reasonability of FT-PCA. The nature of
the second-order-statistics approach is to find a subset of the domain
(either time domain or frequency domain) with positive Lebesgue measure
where the source signals are clearly different. Since the linear combination
matrix $A$ is applied to the whole domain, the BSS problem can be
solved algebraically by finding a subset where the source signals
have the characteristics to be separated. And the reason to pick frequency
domain as the approach is clear: in practice, it makes sense that
different source signals almost always have different density distributions,
and it is almost always possible to find subsets (no matter how small
it is) where the spectrums are approximately exclusive. On the other
hand, using other possible assumptions is less practical, for example,
trying to find an interval in time domain where the signals have exclusive
subsets is unlikely, and thus these kinds of approaches do not work. 

\begin{figure}
\begin{centering}
\includegraphics[scale=0.4]{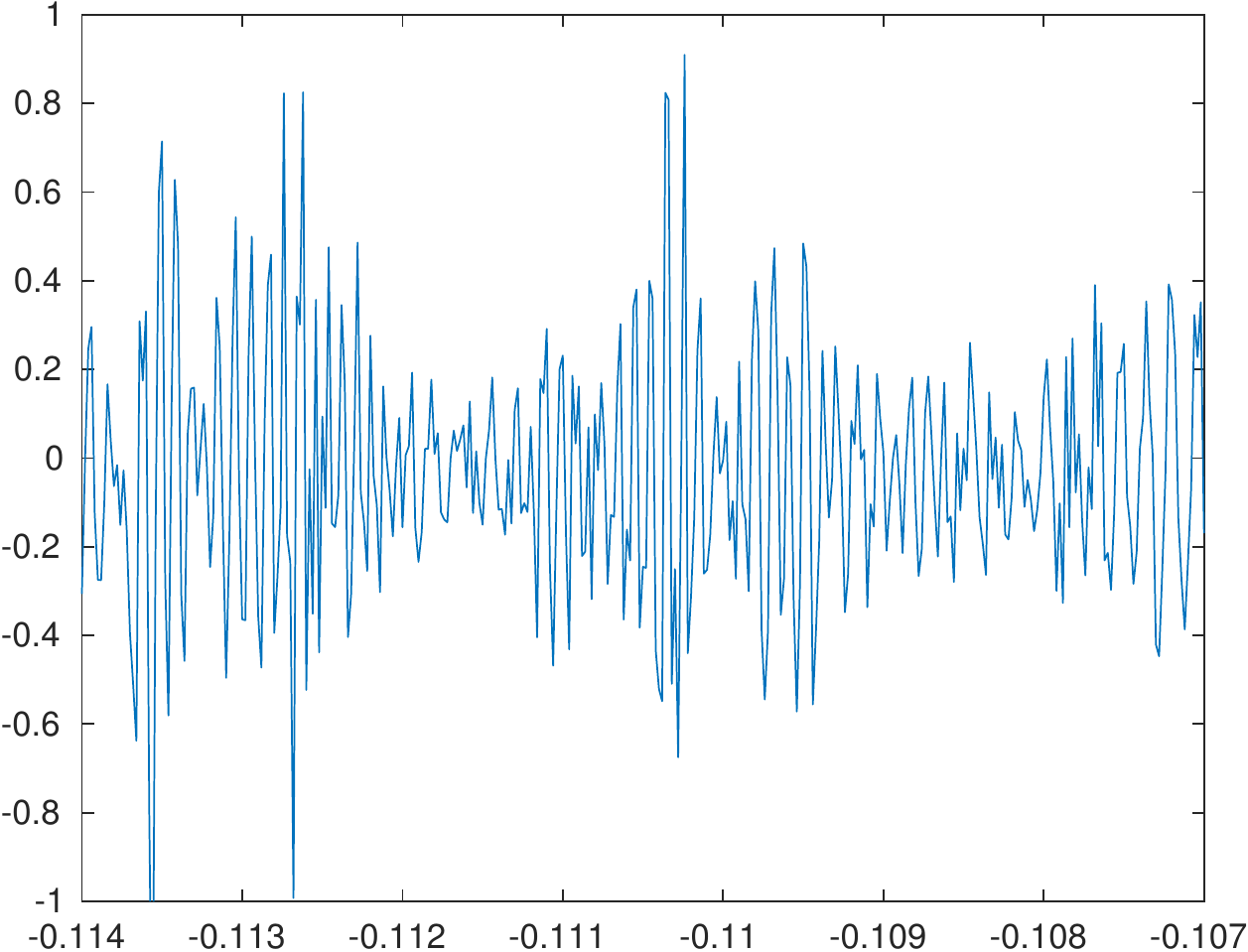}\hspace{2cm}\includegraphics[scale=0.4]{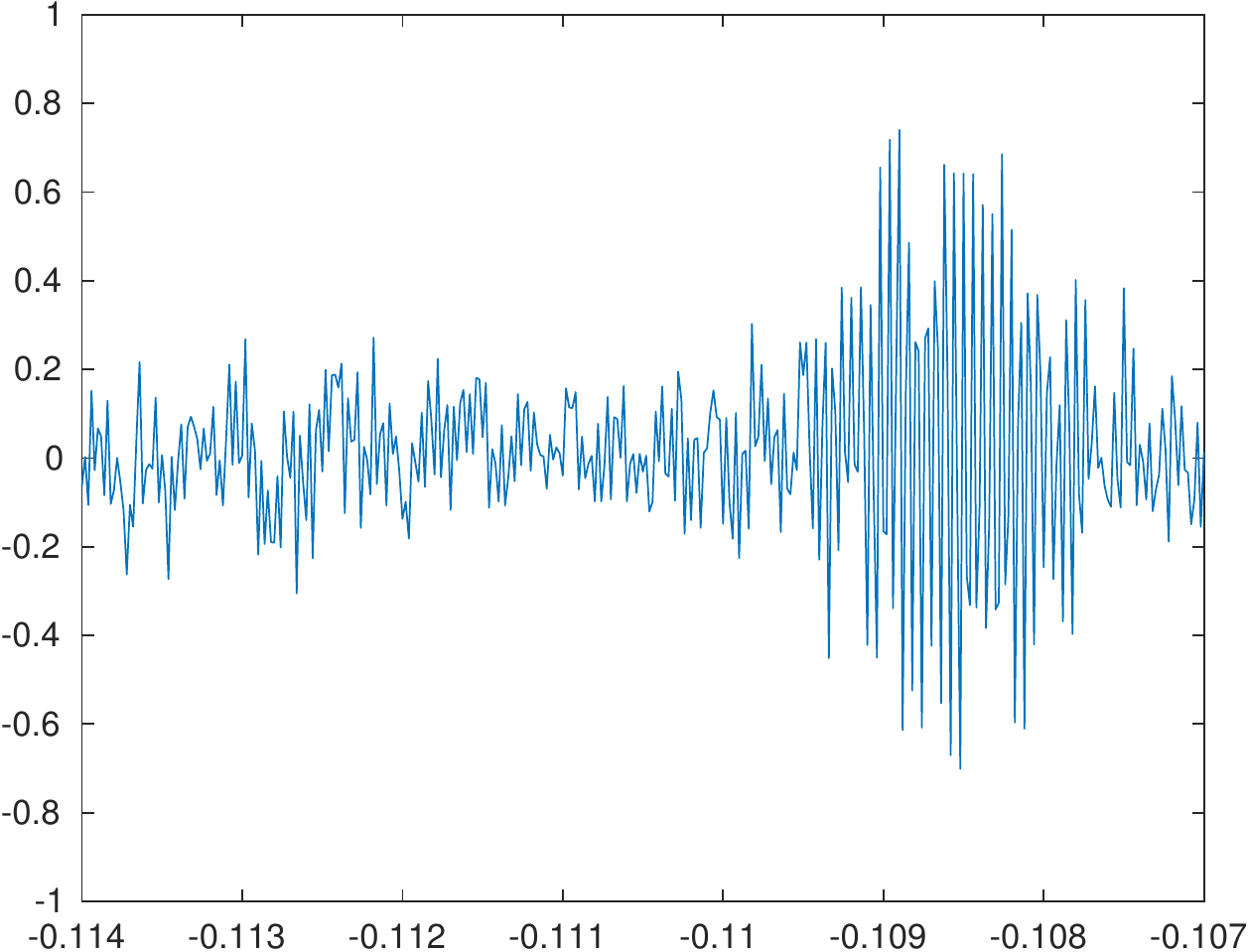}
\par\end{centering}
\caption[An example of an interval where two source signals has approximately
exclusive spectral density. ]{ An example of an interval where two source signals has approximately
exclusive spectral density. The figures show the segments of the Fourier
transform of each signal in this specific interval (only the real
parts). The variance of the left signal in this interval is 0.1626;
the variance of the right signal in this interval is 0.0980l and the
covariance of them is -0.0036+0.0017i.}

\label{fig-omega0exists}
\end{figure}

Practically, since no spectral functions contain subsets where the
spectral power is exactly zero, within acceptable error, if there
exists an interval on which one density function has large values
while the other is close to zero, and vise versa, then the above conditions
can be approximately satisfied. See Fig. \ref{fig-omega0exists}.
And in practice we do not use a true window function as the kernel
but the Eq. \ref{eq:kernel}, in order to make the approximation more
smooth.

\begin{figure*}
\begin{centering}
\includegraphics[scale=0.3]{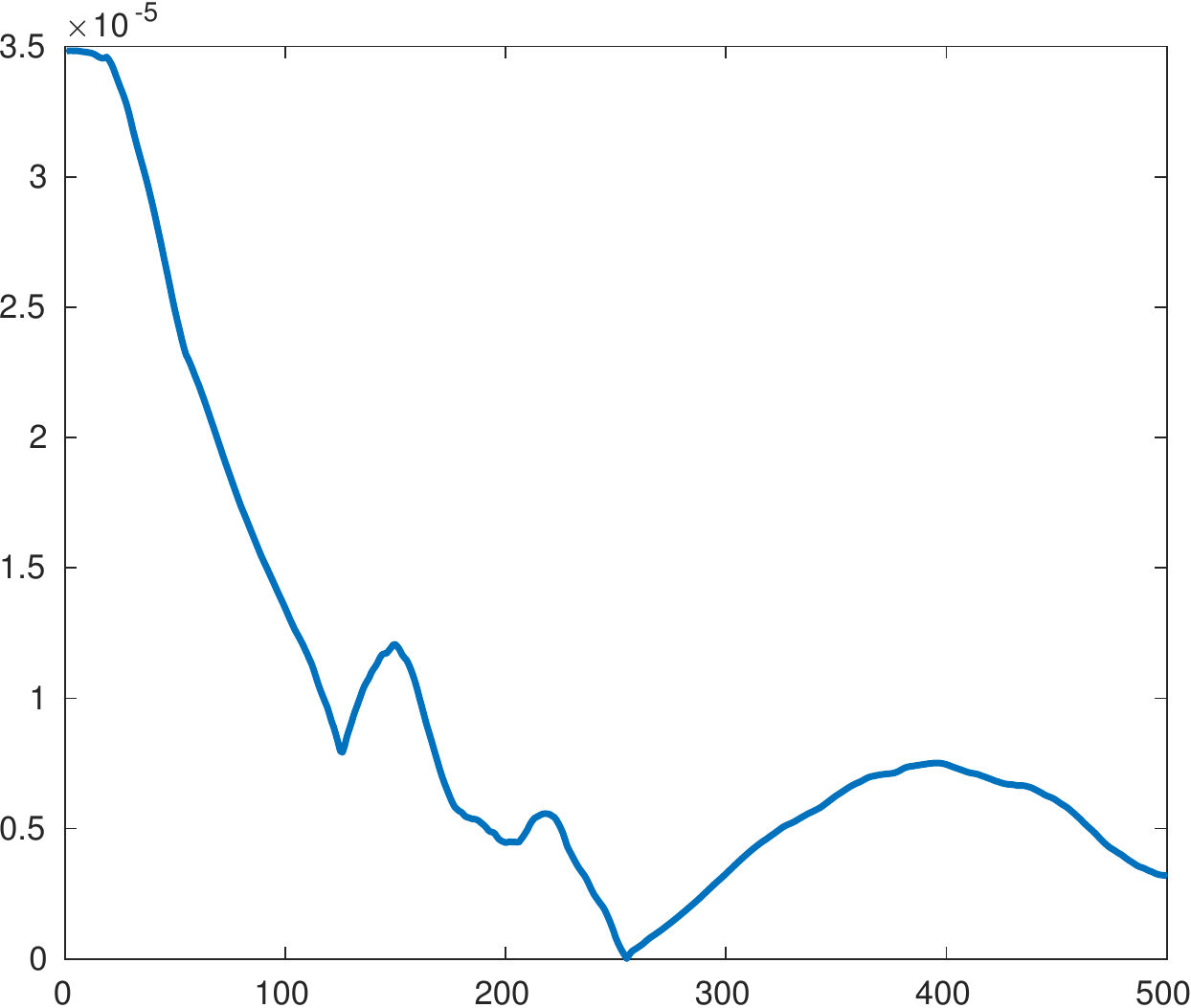}\hspace{0.3cm}\includegraphics[scale=0.3]{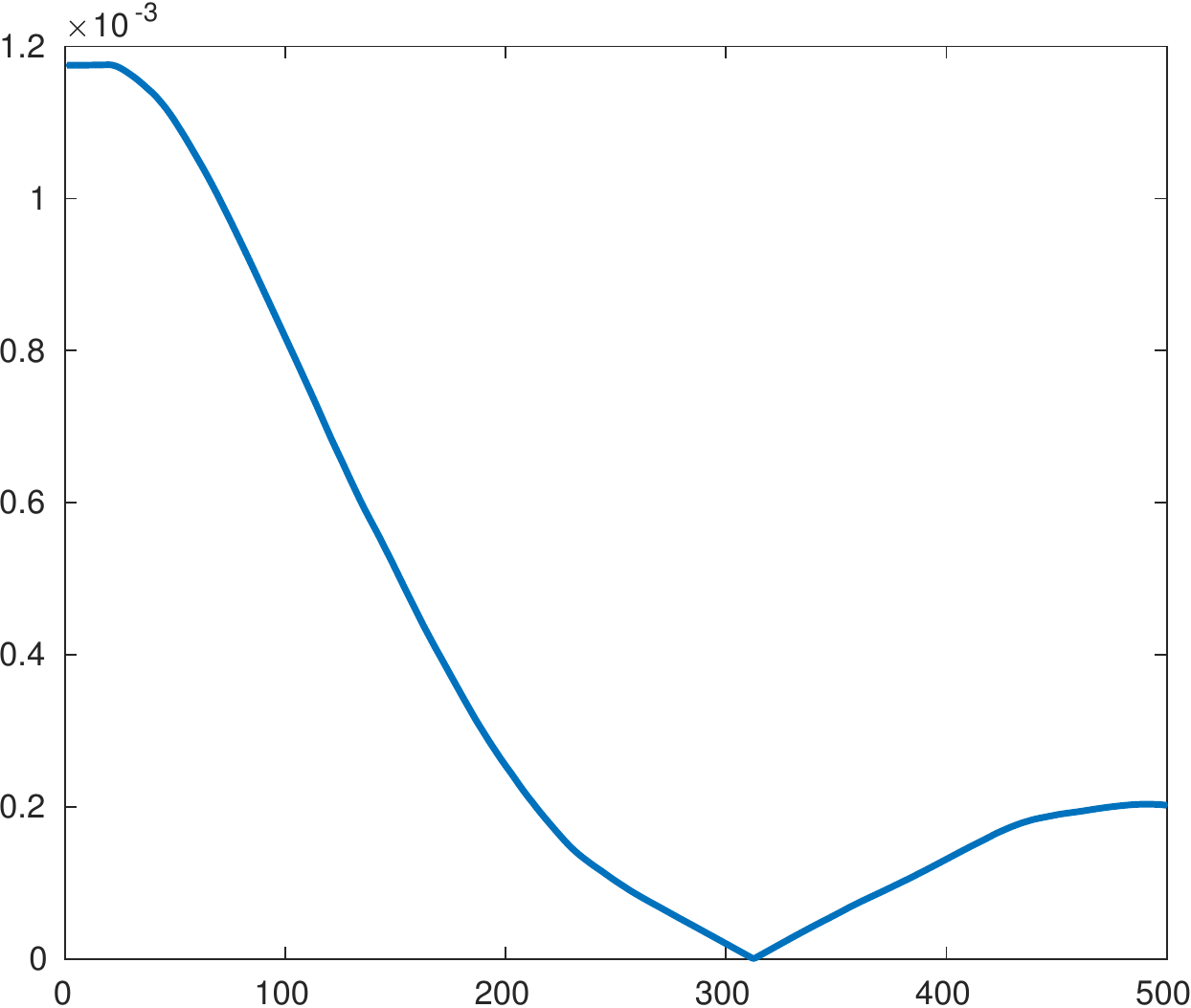}\hspace{0.3cm}\includegraphics[scale=0.32]{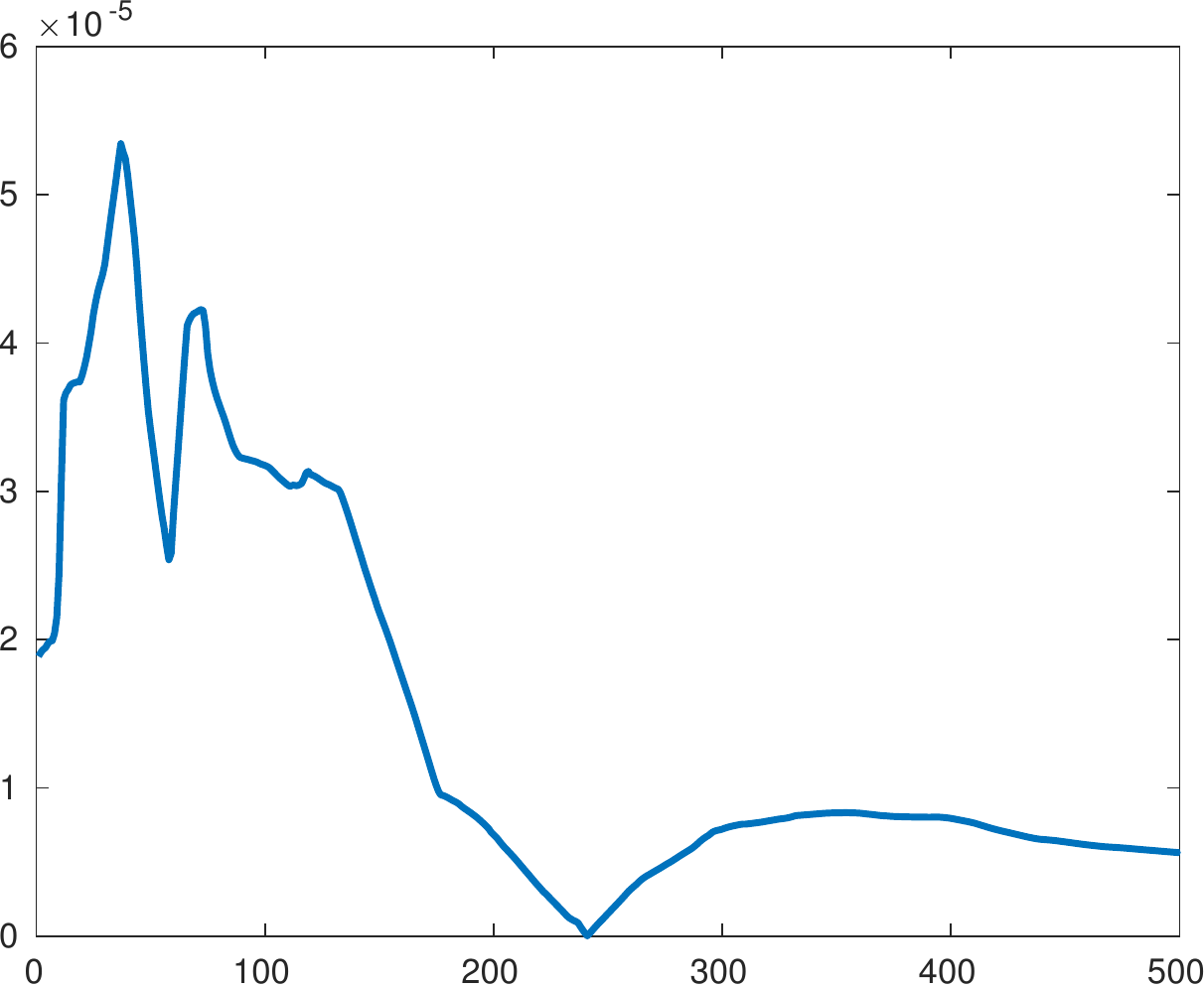}\hspace{0.3cm}\includegraphics[scale=0.32]{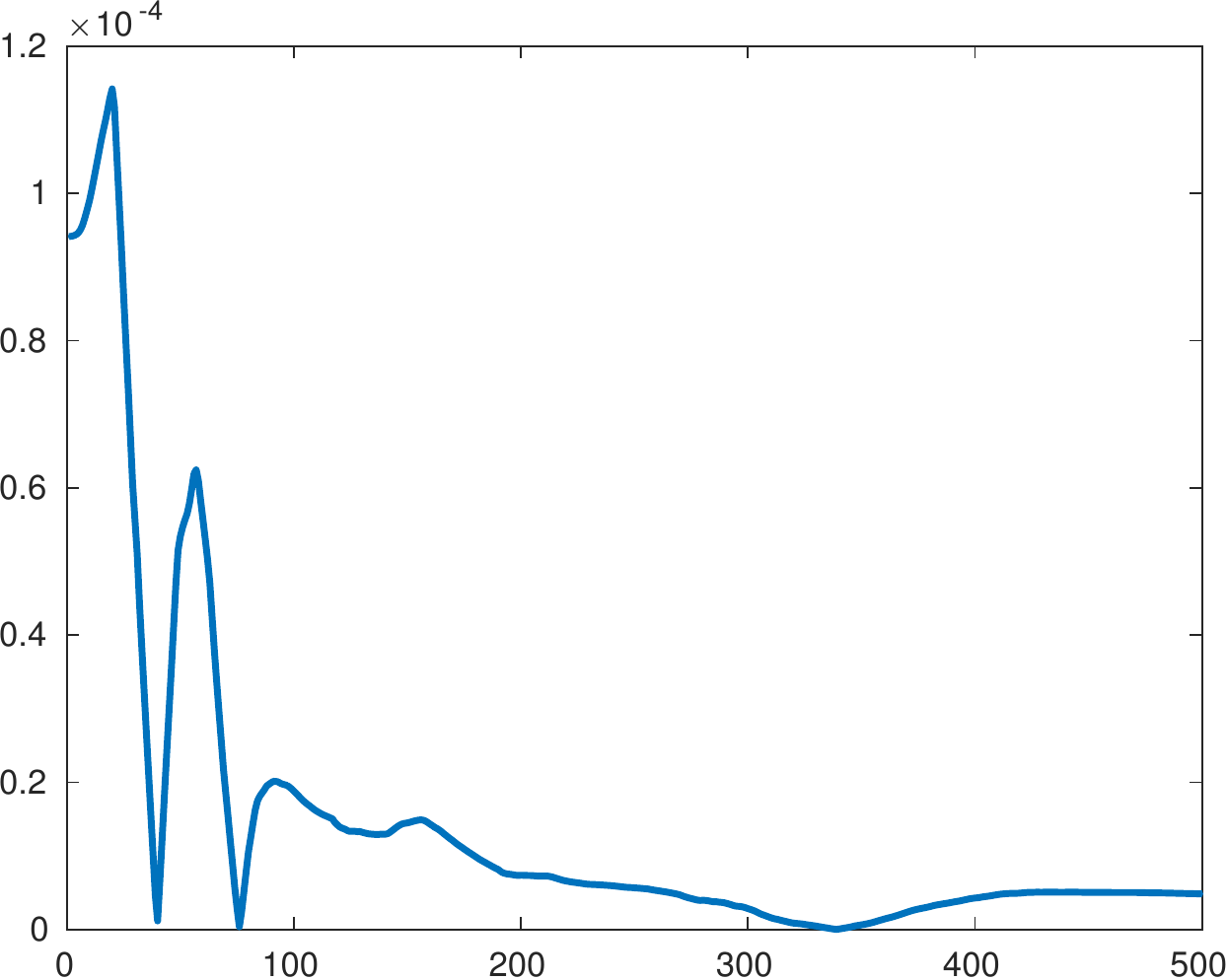}
\par\end{centering}
\caption[Examples of best $\omega_{0}$'s that diagonalize $\int K(\omega)\mathbf{S}(\omega)(\mathbf{S}(\omega))^{H}\mathrm{d}\omega$.]{Examples of best $\omega_{0}$'s that diagonalize $\int K(\omega)\mathbf{S}(\omega)(\mathbf{S}(\omega))^{H}\mathrm{d}\omega$.
Each figure shows the $\mathrm{Re}(\mathbf{S}_{12})$ with respect
to $\omega_{0}$ for a different pair of input source signals. In
each example, there always exists a best $\omega_{0}$ so that $\mathrm{Re}(\mathbf{S}_{12})$
is close to zero the most. With these best $\omega_{0}$'s, the FT-PCA
result has errors (from left to right): 0.000130, 0.0133, 0.00028,
0.00003, which are much smaller than any other approaches. Please
note that, these $\omega_{0}$'s are the ideal cases, based on the
analysis of Section \ref{subsec:The-Heuristic-Strategy}, we know
that there do not exist approaches to search for these ideal $\omega_{0}$'s.
We can only use heuristic strategies to searching for good $\omega_{0}$'s
which has larger errors than these perfect solutions. This figure
is shown to support the reasonability of the assumption of FT-PCA.}

\label{fig:w0_example}
\end{figure*}

On the other hand, from the experimental point of view, we observe
that, for each pair of source signals we examined, there always exists
a best $\omega_{0}$ such that $\int K(\omega)\mathbf{S}(\omega)(\mathbf{S}(\omega))^{H}\mathrm{d}\omega$
is close to a nontrivial diagonal matrix most. And suppose that we
know this specific $\omega_{0}$ for this pair of source signals,
we are able to solve the BSS nearly perfectly using FT-PCA, where
the error is extremely small. See Fig. \ref{fig:w0_example}. This
also supports the reasonability of the assumption of FT-PCA.

\subsection{The Heuristic Strategy to Search for $\omega_{0}$\label{subsec:The-Heuristic-Strategy}}

Unfortunately, though FT-PCA has reasonable assumptions and solid
theory, practically it is not easy to search for the ideal interval
where the source signals are exclusive only based on the input signals
$\mathbf{z}(t)$. This means that for the searching of $\omega_{0}$,
there are no theory to guarantee the optimization. See the following
analysis:

Our task is to apply different $\omega_{0}$ as the shifts of the
kernel function in 
\[
K(\omega)=\frac{1}{1+|\omega-\omega_{0}|},
\]
and search for best $\omega_{0}$ so that $|\mathbf{S}_{11}-\mathbf{S}_{22}|$
is not close to zero while $|\mathbf{S}_{12}|$ is minimized, based
on the values of $\mathbf{Z}_{ij}$ for $i,j=1,2$ that we computed
according to each $\omega_{0}$ that we apply. Practically, we need
$\int K(\omega)\mathbf{S}(\omega)(\mathbf{S}(\omega))^{H}\mathrm{d}\omega$
more ``diagonal'' than $\int K(\omega)\mathbf{Z}(\omega)(\mathbf{Z}(\omega))^{H}\mathrm{d}\omega$.

Without loss of generality, we can write the rotation matrix 
\[
\hat{R}=\left(\begin{array}{cc}
\cos\theta & -\sin\theta\\
\sin\theta & \cos\theta
\end{array}\right)
\]
for a certain rotation angle $\theta$. Then, by\textit{
\[
\int K(\omega)\mathbf{Z}(\omega)(\mathbf{Z}(\omega))^{H}\mathrm{d}\omega=\hat{R}^{T}\int K(\omega)\mathbf{S}(\omega)(\mathbf{S}(\omega))^{H}\mathrm{d}\omega\hat{R}
\]
} we have\textit{
\[
\mathbf{Z}_{11}=\cos^{2}\theta\mathbf{S}_{11}+2\cos\theta\sin\theta\mathrm{Re}(\mathbf{S}_{12})+\sin^{2}\theta\mathbf{S}_{22}
\]
\begin{equation}
\mathbf{Z}_{12}=\cos\theta\sin\theta(\mathbf{S}_{22}-\mathbf{S}_{11})+(\cos^{2}\theta-\sin^{2}\theta)\mathrm{Re}(\mathbf{S}_{12})+\mathrm{Im}(\mathbf{S}_{12})\label{eq:3-1}
\end{equation}
\[
\mathbf{Z}_{21}=\cos\theta\sin\theta(\mathbf{S}_{22}-\mathbf{S}_{11})+(\cos^{2}\theta-\sin^{2}\theta)\mathrm{Re}(\mathbf{S}_{12})-\mathrm{Im}(\mathbf{S}_{12})
\]
\[
\mathbf{Z}_{22}=\sin^{2}\theta\mathbf{S}_{11}-2\cos\theta\sin\theta\mathrm{Re}(\mathbf{S}_{12})+\cos^{2}\theta\mathbf{S}_{22}.
\]
}

Clearly, each $\mathbf{Z}_{ij}$ is a mixture of $\mathbf{S}_{11},\mathbf{S}_{22}$,
and $\mathbf{S}_{12}$, and $|\mathbf{S}_{11}-\mathbf{S}_{22}|$ and
$|\mathbf{S}_{12}|$ cannot be solved separately by $\mathbf{Z}_{ij}$
without knowing $\theta$. And we are not able to understand the changing
of $|\mathbf{S}_{11}-\mathbf{S}_{22}|$ and $|\mathbf{S}_{12}|$ by
observing the changing of $\mathbf{Z}_{ij}$, either. Hence, theoretically
there is no way to guarantee that the optimized $\omega_{0}$ can
be searched based on the $\mathbf{Z}_{ij}$ values we observed.

However, there exist heuristic strategies to search for good $\omega_{0}$. 

From the eigen decomposition structure of the equation
\[
\int K(\omega)\mathbf{Z}(\omega)(\mathbf{Z}(\omega))^{H}\mathrm{d}\omega=\hat{R}^{T}\int K(\omega)\mathbf{S}(\omega)(\mathbf{S}(\omega))^{H}\mathrm{d}\omega\hat{R}
\]
and the relationship of traces and determinants, we observe that:
\begin{equation}
\mathbf{Z}_{11}+\mathbf{Z}_{22}=\mathbf{S}_{11}+\mathbf{S}_{22}\label{eq:1}
\end{equation}
and
\begin{equation}
\mathbf{Z}_{11}\mathbf{Z}_{22}-|\mathbf{Z}_{12}|^{2}=\mathbf{S}_{11}\mathbf{S}_{22}-|\mathbf{S}_{12}|^{2}.\label{eq:2-1}
\end{equation}

\begin{figure*}
\begin{centering}
\includegraphics[scale=0.35]{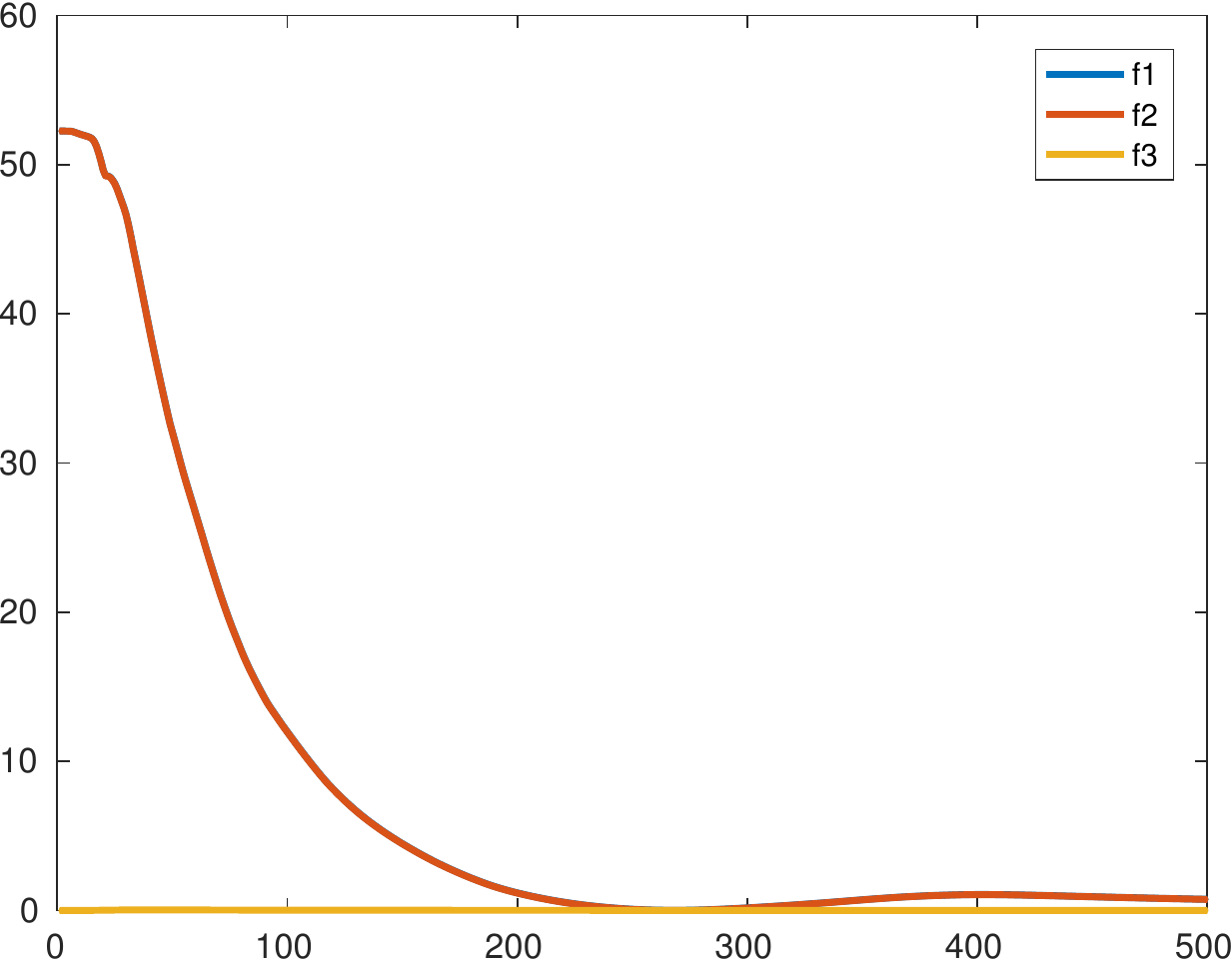}\hspace{1cm}\includegraphics[scale=0.35]{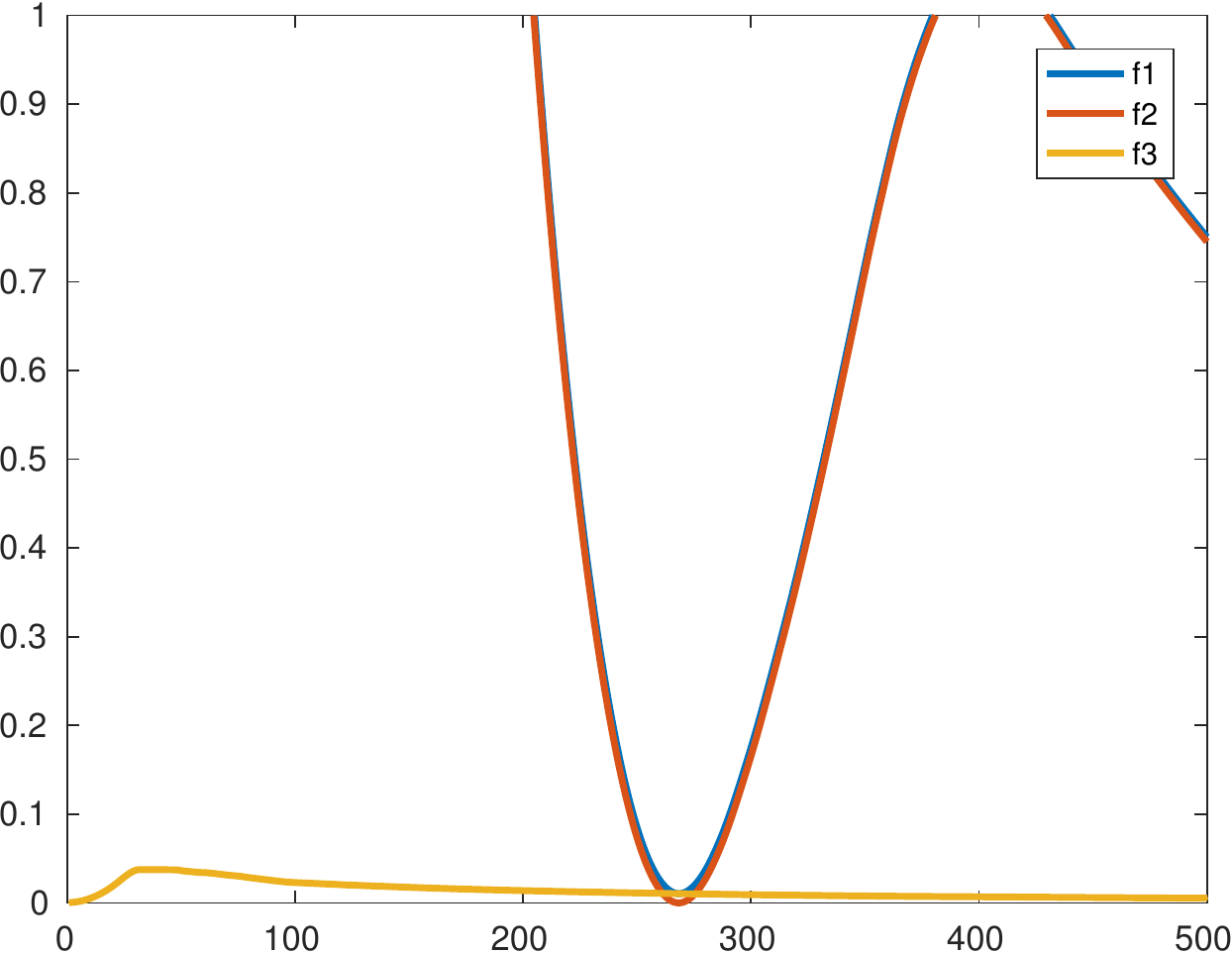}\hspace{1cm}\includegraphics[scale=0.35]{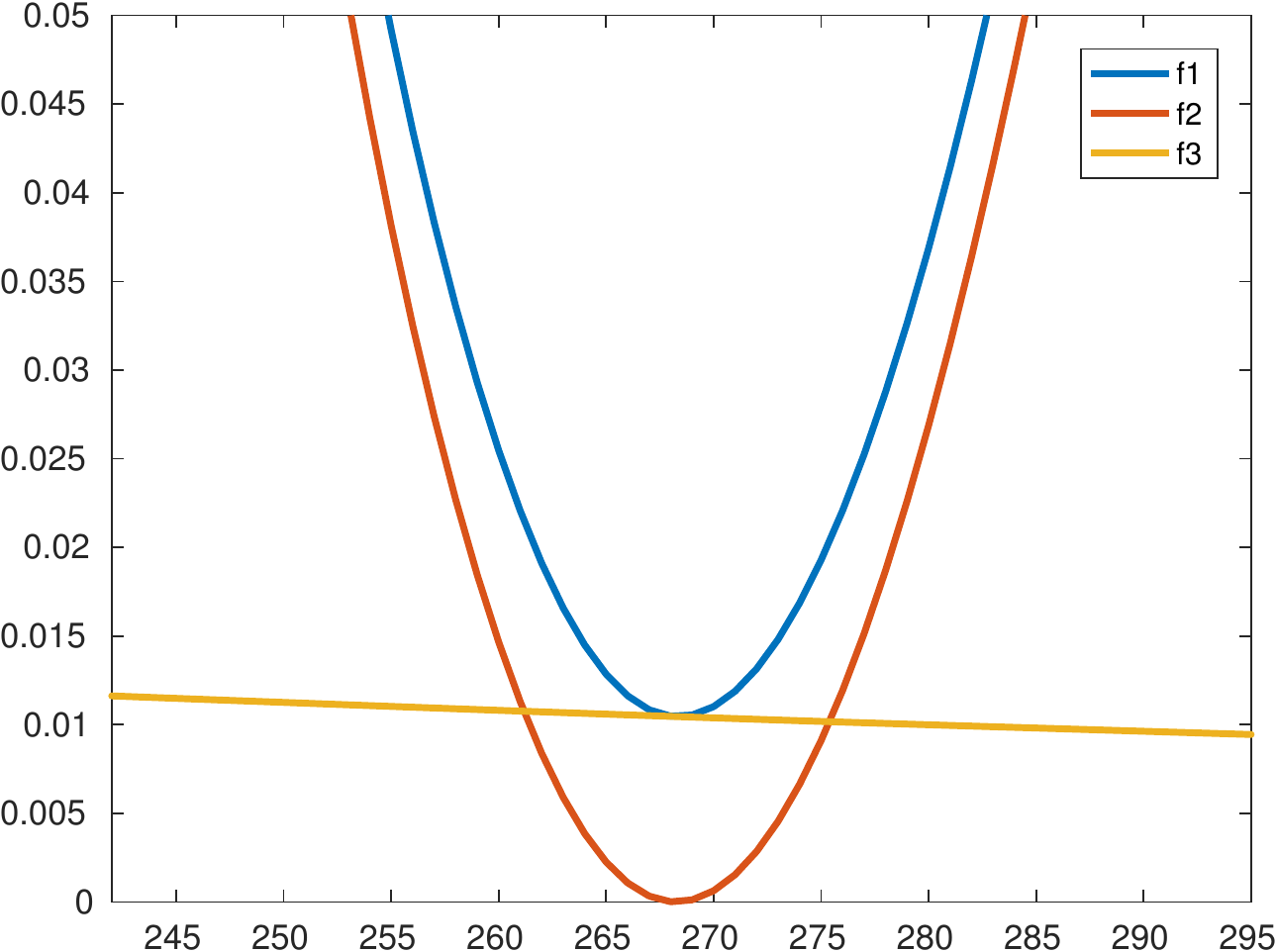}
\par\end{centering}
\caption[The function graphs of $f_{1},f_{2},f_{3}$ with respect to the kernel
shift $\omega_{0}$.]{The function graphs of $f_{1},f_{2},f_{3}$ with respect to the kernel
shift $\omega_{0}$. \textbf{Left:} The overall function graphs.
\textbf{Mid: }A closer view by showing the function values within
range $[0,1]$. \textbf{Right: }The function graphs around the global
minimum, where $f_{2}$ falls below $f_{3}$. The good $\omega_{0}$
can be selected at the points outside the crossing of $f_{2}$ and
$f_{3}$, where $f_{3}$ is small enough and $f_{2}$ is not close
to zero.}

\label{fig.6-heuristic_strategy}
\end{figure*}

From Eq. \ref{eq:1} and \ref{eq:2-1}, we have
\[
(\mathbf{Z}_{11}-\mathbf{Z}_{12})^{2}+4|\mathbf{Z}_{12}|^{2}=(\mathbf{S}_{11}-\mathbf{S}_{22})^{2}+4|\mathbf{S}_{12}|^{2}
\]
where the left hand side is available, while the right hand side contains
the sum of squares of the two terms that we care about most. For convenience,
let's name them as follows:
\[
f_{1}=(\mathbf{Z}_{11}-\mathbf{Z}_{12})^{2}+4\mathbf{Z}_{12}^{2}
\]
\[
f_{2}=(\mathbf{S}_{11}-\mathbf{S}_{22})^{2}
\]
\[
f_{3}=4|\mathbf{S}_{12}|^{2}.
\]
Since $f_{3}$ is close to zero, the values of $f_{1}$ is dominated
by $f_{2}$. However, maximizing $f_{1}$ is not a good strategy,
since we do not need $F_{2}$ to be maximized but just not close to
zero, and practically by maximizing $f_{1}$, the part of $f_{3}$
gets larger which leads to worse solution, since the diagonal property
of $\int K(\omega)\mathbf{S}(\omega)(\mathbf{S}(\omega))^{H}\mathrm{d}\omega$
is more sensitive with the changing of $f_{3}$. From Fig. \ref{fig.6-heuristic_strategy}
we can have an intuitive idea about the absolute values of $f_{1},f_{2},f_{3}$.
From the figure, as well as observations on other examples, we notice
that, minimizing instead of maximizing $f_{1}$ should be a better
solution, since the magnitude of $f_{3}$ will be controlled while
$f_{1}$ decreases, which can guarantee $f_{3}$ being close to zero.
After the minimum of $f_{1}$ is found, we need to move the shift
$\omega_{0}$ steps away from the $\arg\min f_{1}$. This is because
that there exists a small interval around $\arg\min f_{1}$ where
$f_{2}$ drops heavily so as to be even less than $f_{3}$, and practically
this will lead to $\int K(\omega)\mathbf{S}(\omega)(\mathbf{S}(\omega))^{H}\mathrm{d}\omega$
getting too close to the identity matrix. By moving away a certain
distance from the minimum point, we are able to have $f_{2}$ significantly
larger than $f_{3}$, and practically this can be considered as $|\mathbf{S}_{11}-\mathbf{S}_{22}|$
being far from zero while $|\mathbf{S}_{12}|$ is close to zero. From
the experient results shown in Section \ref{sec:6-Experiments}, through
this heuristic strategy we were able to get notable results even better
than SOBI.

\section{Experiments\label{sec:6-Experiments}}

\subsection{The Behaviors of the Geometrical Objective Functions\label{subsec:The-Behaviors-of}}

\begin{figure}
\begin{centering}
\includegraphics[scale=0.25]{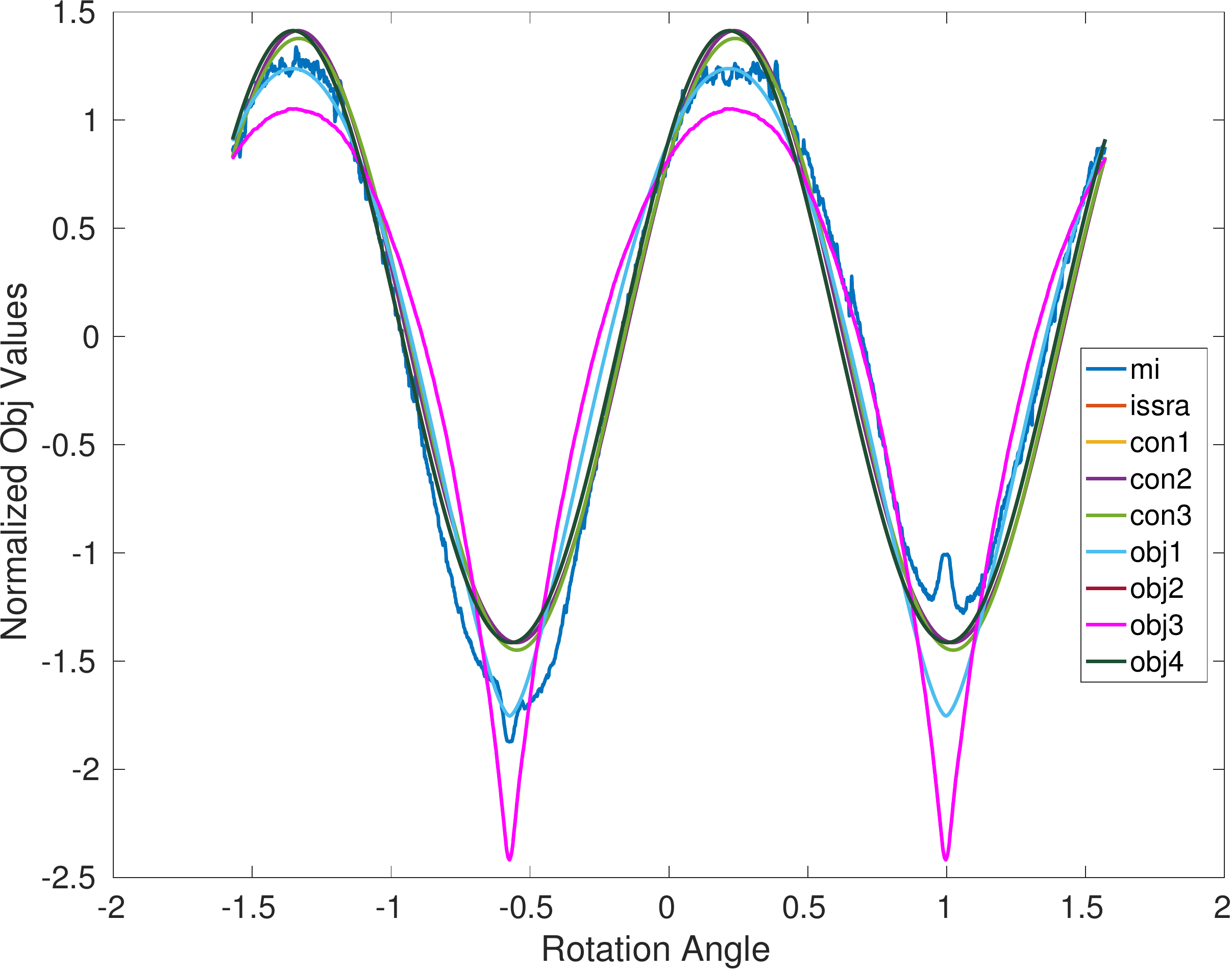}\hspace{2cm}\includegraphics[scale=0.26]{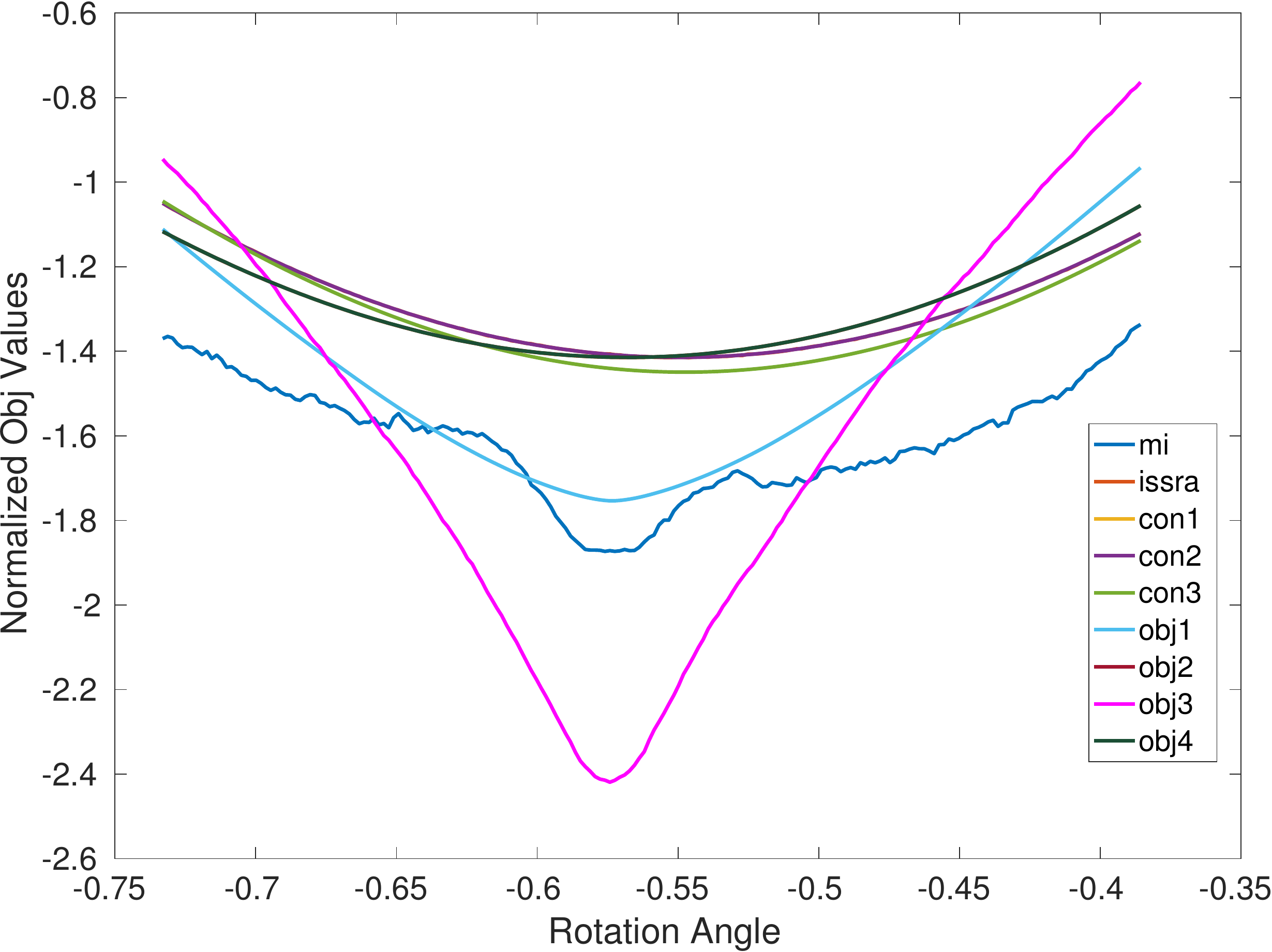}
\par\end{centering}
\caption[Comparison of different objective functions for the second step.]{Comparison of different objective functions for the second step.
\textbf{Left}: The overall objective function graphs. \textbf{Right}:
The objective function graphs in a neighborhood of the global minimum.}

\label{fig6-1}
\end{figure}

Firstly, we show the function graphs of different objective functions
proposed in Section \ref{subsec:Geometrical-Objective-functions},
and compare them with MI and contrast functions proposed for FastICA.
We randomly picked two natural signals (two segments of true audio
files) $s_{1}(t)$ and $s_{2}(t)$, normalized them by removing their
mean and standardizing their covariance matrix, and then applied a
random $2\times2$ mixing matrix $A$ to get the mixed signal observations
$x_{1}(t)$ and $x_{2}(t)$. Then the first step \textendash{} PCA
was operated on the observed signals to get the standardized signals
$z_{1}(t)$ and $z_{2}(t)$. Finally, we searched the angle $\theta$
from $-\pi$ to $\pi$, and computed the recovered signals $y_{1}(t)$
and $y_{2}(t)$ by

\[
\mathbf{y}_{\theta}(t)=R(\theta)\mathbf{z}(t)
\]
where $R(\theta)$ is the rotation matrix
\[
\left(\begin{array}{cc}
\cos\theta & -\sin\theta\\
\sin\theta & \cos\theta
\end{array}\right)
\]
of each rotation angle $\theta$, and computed the values of each
objective functions, i.e. $\mathrm{Obj}(\mathbf{y_{\theta}}(t))$
. Fig. \ref{fig6-1} shows the objective function graphs with respect
to the rotation angle. The objective values shown in the figure were
normalized (removing means and divided by standard deviations) so
that they are comparable. In the figures, 'mi' means the original
MI objective, i.e. the sum of marginal entropies; 'issra' is the objective
function $\mathcal{O}_{1}$ in Section \ref{subsec:Geometrical-Objective-functions};
'con1' to 'con3' are the FastICA contrast functions $G_{1}$ to $G_{3}$;
and 'obj1' is $\mathcal{O}_{3}$, 'obj2' is $\mathcal{O}_{2}$, 'obj3'
is $\mathcal{O}_{4}$, and 'obj4' is $\mathcal{O}_{5}$, in Section
\ref{subsec:Geometrical-Objective-functions}.

From Fig. \ref{fig6-1}, we observe that for all the objective functions,
their function graph has similar shapes, and the global minimum are
very close. This shows the fact that all the proposed geometrical
objective functions are good approximations to the MI objective (which
is in fact the summation of marginal entropies). 

We also did synthetic experiments to investigate the behaviors of
each geometrical objective functions, as well as the MI objective
and FastICA contrast functions. The synthetic experiment was done
the same way as above: Firstly we randomly picked a pair of source
signals, and standardize them. Then we applied random mixing matrix
to generate the observed signals. For the BSS process, we did the
PCA-ICA steps, and in the ICA step, we optimized each objective function
to get its solution, as well as the computation time. The synthetic
experiment was repeated for 20 times, and the errors and CPU time
were provided as mean/std of the 20 results for each objective function.

Table \ref{tab6-1} shows the differences between the MI results and
each other objective function, where $\mathcal{\tilde{O}}_{3}=|\int\prod_{i}y_{i}'(t)\mathrm{d}t|$
is the objective function of the Derivative-PCA. The errors were computed
as the differences of the resulted rotation angles. From the table,
we observe that, $\mathcal{O}_{3}$ approximates the MI objective
function best. 

\begin{table*}
\caption{Average error of each objective function with MI.}

\begin{centering}
\begin{tabular}{|c|c|c|c|c|c|c|c|c|c|}
\hline 
 & $\mathcal{O}_{1}$ & $\mathcal{O}_{2}$ & $\mathcal{O}_{3}$ & $\mathcal{O}_{4}$ & $\mathcal{O}_{5}$ & $\mathcal{\tilde{O}}_{3}$ & $G_{1}$ & $G_{2}$ & $G_{3}$\tabularnewline
\hline 
\hline 
Error (rad) & 0.0623 & 0.0623 & \textbf{0.0621} & 0.4484 & 0.0624 & 0.0680 & 0.3457 & 0.0706 & 0.3463\tabularnewline
\hline 
\end{tabular}
\par\end{centering}
\label{tab6-1}
\end{table*}

Table \ref{tab6-2} shows the mean errors computed between the solution
of each objective function and the ground truth. The errors were the
differences between the resulted rotation angles of each objective
function and the true rotation angle. Please note that, since we standardized
the source signals, the source signals and the first step (PCA) results
exactly differ by a rotation, and hence we can compare the true rotation
angle with each approach. The optimizations were done by brute-force
search. The CPU time was the average of all 20 repeated experiments.
In each experiment, each objective function was computed 1801 times
(from $-180^{\circ}$ to $180^{\circ}$ with step size as $0.1^{\circ}$).

\begin{table*}
\caption{Average error of each objective function with the ground truths and
average computational time of each objective functions.}

\begin{centering}
\begin{tabular}{|c|c|c|c|c|c|c|c|c|c|}
\hline 
 & MI & $\mathcal{O}_{1}$ & $\mathcal{O}_{2}$ & $\mathcal{O}_{3}$ & $\mathcal{O}_{5}$ & $\mathcal{\tilde{O}}_{3}$ & $G_{1}$ & $G_{2}$ & $G_{3}$\tabularnewline
\hline 
\hline 
Error (deg) & 0.7683 & 0.7685 & 0.7735 & \textbf{0.7535} & \textbf{0.7535} & 4.8035 & 2.2080 & 1.4273 & 1.3230\tabularnewline
\hline 
Std & 1.0697 & 1.0679 & 1.0840 & 1.0668 & 1.0703 & 2.1372 & 4.3855 & 1.1455 & 1.1539\tabularnewline
\hline 
Time (sec) & 6.14 & 10.40 & 9.55 & \textbf{2.13} & 2.62 & 2.14 & 3.00 & \textbf{1.39} & 7.33\tabularnewline
\hline 
\end{tabular}
\par\end{centering}
\label{tab6-2}
\end{table*}

From the table we can observe that all the geometrical objective functions
worked well, much better than all contrast functions. And the best
ones: $\mathcal{O}_{3}$ and $\mathcal{O}_{5}$ worked even slightly
better than MI. Among the geometrical objectives, $\mathcal{O}_{3}$
has the least computational time due to its simple formula. The experiment
results supports our assertion that the geometrical objective functions
are good candidates for the ICA step of BSS problems, especially $\mathcal{O}_{3}$
and $\mathcal{O}_{5}$ which compute simply and fast, and has promisingly
good precision for mixed signal recovering. Additionally, being preliminary
convex functions with their derivatives available, $\mathcal{O}_{3}$
and $\mathcal{O}_{5}$ can be applied to a gradient-based optimization
algorithm, and serve as a good algorithm for ICA-based BSS, replacing
the traditional MI objective functions. The only disadvantage of $\mathcal{O}_{3}$
is that the time complexity gets higher than FastICA as the number
of signals grows. And $\mathcal{O}_{5}$ can be adapted to the FastICA
algorithm for it has the form of aggregation.

\subsection{The Comparison of the Geometrical Objective Functions and the FT-ICA
approach}

In this experiment, we compare the MI objective with the above proposed
objective $\mathcal{O}_{3}$, as well as the contrast function $G_{2}$
(which approximates MI well and has fastest computational time in
the above experiments) and the second-order approaches: SOBI and FT-PCA.
The inputs are 72 pairs of real source signals (audio segments) that
were standardized to have zero mean and identity covariance. Random
mixing matrices were applied to each pair of source signals, and the
compared approaches were applied to solve for the second rotation.
The errors were computed as the differences of the rotation angles
solved by each approach with the ground truth. Table \ref{tab6-3}
shows the average error for all 72 rounds of experiments, without
noise or signal-noise-ratio (SNR) being 100, 50, and 20. For the FT-PCA
approach, we adopted the heuristic strategy that we described in Section
\ref{subsec:The-Heuristic-Strategy}, where the searching radius after
minimization of $f_{1}$ is fixed as 100 and the step size of $\omega_{0}$
is 0.001.

\begin{table*}
\caption{The error table of different compared approaches for the synthetic
BSS experiment.}

\begin{centering}
\begin{tabular}{|c|c|c|c|c|c|c|}
\hline 
Error (deg) & MI & $G_{2}$ & $\mathcal{O}_{3}$ & FT-PCA & Derivative-PCA & SOBI\tabularnewline
\hline 
\hline 
No Noise & \textbf{0.7668} & 1.0995 & 0.7807 & \textbf{\textcolor{blue}{0.8735}} & \textcolor{blue}{2.7195} & \textcolor{blue}{1.0195}\tabularnewline
\hline 
SNR = 100 & 0.7970 & 1.1447 & \textbf{0.7878} & \textbf{\textcolor{blue}{0.8158}} & \textcolor{blue}{2.8338} & \textcolor{blue}{0.9767}\tabularnewline
\hline 
SNR = 50 & 0.8777 & 1.1796 & \textbf{0.8724} & \textbf{\textcolor{blue}{1.0363}} & \textcolor{blue}{2.9054} & \textcolor{blue}{1.0750}\tabularnewline
\hline 
SNR = 20 & 8.026 & 5.612 & 5.779 & \textcolor{blue}{23.856} & \textcolor{blue}{24.253} & \textcolor{blue}{17.942}\tabularnewline
\hline 
\end{tabular}
\par\end{centering}
\label{tab6-3}
\end{table*}

From the table, we observe that, when there are no noise, or the SNR
= 100 and 50, every approach works well. Errors of all approaches
except Derivative-PCA are less than 2 degree, which indicates that
all these approaches have practically acceptable precision. For the
optimization approaches (MI, $G_{2}$, and $\mathcal{O}_{3}$), MI
and $\mathcal{O}_{3}$ worked better, and $G_{2}$ worked worse. And
MI is slightly worse than $\mathcal{O}_{3}$ in average though they
are very close. For the second-order-statistics approaches, FT-PCA
worked best (its precision is very close to MI), and Derivative-PCA
worked worst. When the SNR is 20, none of these approaches worked.
The above results supports that, for optimization based approaches,
$\mathcal{O}_{3}$ is indeed a good objective function, which is significantly
better than the contrast functions and competitive with MI. For second-order-statistics
approaches, FT-PCA works better than SOBI, hence it is an effective
approach for BSS problems that is based on different assumptions than
independence and has simple and fast algorithm.
\begin{singlespace}

\section{Conclusions\label{sec:Conclusions}}
\end{singlespace}

In this paper, we highlight two main contributions. First, we point
out the model of ICA-based approaches for BSS, and based on the relationship
between SPC and MI, we apply SPC to BSS to propose geometrical objective
functions based on the property of the joint signals, whose computational
time and precision are both excellent. Second, we proposed a new second-order-statistics
approach, FT-PCA, that assumes the kernel orthogonality of signals
in the frequency domain, and solve the BSS problem by applying Fourier
transforms and solve a second eigen decomposition. Comparing with
other second-order-statistics approaches, FT-PCA has a more reasonable
assumption that bypasses the independence concepts, and has a simple
and fast algorithm that does not require any optimization or joint
diagonalization, given good hyper-parameters. We also propose heuristic
strategies for searching good hyper-parameter, which was proven efficient
in the experiment section.

A immediate future work is to extend the idea of FT-PCA to nonstationary
signals, and propose a generalized algorithm that works for signals
which have different frequency distribution for different time intervals.
Another potential future work is to apply FT-PCA to nonlinear ICA
problems.

\section*{Acknowledgments}

We acknowledge helpful conversations with Mingyuan Gao and Yuan Zhou.

\vspace{-5pt}

\bibliographystyle{abbrv}
\bibliography{/home/douglas/Dropbox/Research/Aug2017_ICA/reference}

\end{document}